\documentclass[twocolumn,
showpacs,preprintnumbers,amsmath,amssymb,floatfix,aps]{revtex4}

\usepackage[dvips]{graphicx}
\usepackage{dcolumn}

\begin{document}
\title{DYNAMICS AND SUPERFLUIDITY OF AN ULTRACOLD FERMI GAS}

\author{Sandro Stringari}
\affiliation{Dipartimento di Fisica, Universita' di Trento and CNR-INFM
BEC Center, 1-38050 Povo, Italy}
\maketitle

\section{Introduction}

After the first experimental realization of Bose-Einstein condensation in dilute atomic gases \cite{BEC} the field of ultracold gases has become a rapidly growing field of research (for reviews see, for example, \cite{varenna,RMP,LeggettRMP,RMPNobel,BookDK,BookTN}.
In the last years a considerable amount of experimental and theoretical work has focused on Fermi gases. With respect to Bose gases Fermi systems exhibit important differences which are the consequence of quantum statistics and of the role of  interactions. A first important difference is that, at low temperature, dilute Fermi gases occupying a single spin state practically do not interact since s-wave scattering is suppressed by  the Pauli principle. This provides  a unique opportunity  for an almost perfect realization of the ideal Fermi gas, with useful applications to precision measurements and quantum information processes. A second important difference is that 
superfluidity in Fermi gases is the result of  non trivial many-body mechanisms where   interactions play a crucial role giving rise to pairing effects. The resulting many-body state exhibits a rich variety of features, depending on the sign and the value of the scattering length characterizing the interaction between atoms belonging to different atomic species. A particularly interesting  configuration is the so called unitary limit  where the scattering length takes an infinite value. At unitarity  the system is found to be particularly stable and to exhibit clean manifestations of superfluidity. 
The possibility of tuning the value of the scattering length profiting of the existence of Feshbach resonances and the rich variety of trapping configurations, both of magnetic and  optical nature, are making the study of ultracold Fermi gases a rich subject of research with many stimulating opportunities from both the experimental and theoretical side.
   
The purpose of this paper is to review some of the dynamic and superfluid  features  exhibited by ultracold Fermi  gases with special emphasis on the effects of the external confinement which will be assumed in most cases of harmonic shape.
 After introducing the main features of the ideal Fermi gas in a harmonic trap (Sect. 2) we will discuss the role of interactions and the general behavior exhibited by an interacting Fermi gas along the BEC-BCS crossover (Sect. 3). We will then apply the many body results of Sect. 3 to trapped configurations (Sect. 4). Sects. 5 and 6 will be devoted to the study of superfluid effects, both concerning the dynamic behavior (expansion and collective oscillations) and the rotational properties (Sect. 6). 
  
\section{Ideal Fermi gas in harmonic trap}

The ideal Fermi gas  represents a natural starting point for discussing the physics of dilute Fermi gases.
In some cases the role of interactions can be in fact neglected (this happens for spin polarized gases where the interaction is strongly suppressed by the antisymmetrization requirement). Viceversa when the role of interactions becomes crucial the comparison with the predictions of the ideal gas  points out explicitly new interesting features exhibited by Fermi gases.

The ideal Fermi gas in the harmonic potential
 \begin{equation}
V_{ho}=\frac{1}{2}m\omega_x^2x^2 + \frac{1}{2}m\omega_y^2y^2 + \frac{1}{2}m\omega_z^2z^2 
\label{Vho}
\end{equation} 
is a
subject with many applications in different fields of physics, ranging from
nuclear physics to quantum dots and is widely discussed in the lterature. For this reason
we will only focus on the most relevant features of the model, emphasizing
the large $N$ behavior where the motion of the gas can be described in
semiclassical terms. In the large $N$ limit many single particle states are in fact occupied and the role of the Heisenberg uncertainty principle can be safely ignored for most properites of the system. 
The simplest way to introduce the semiclassical description is to use a
local density approximation for the Fermi distribution function of a given spin species:
\begin{equation}
f({\bf r},{\bf p}) = \frac{1 }{\exp[\beta \left( p^2/2m + V_{ho}({\bf r})-\mu \right)] +1 }  
\label{fF}
\end{equation}
where $\mu$ is the chemical potential fixed by the normalization condition
\begin{equation}
N_\sigma = \frac{1 }{(2\pi \hbar)^3}\int d{\bf r} d{\bf p} f({\bf r},{\bf p})= 
\int_0^{\infty} {g(E) dE \over \exp[\beta
(E-\mu)] + 1} 
\label{NF}
\end{equation}
and $N_\sigma$ is  the number of atoms of the given spin species. Equation (\ref{fF}) explicitly shows that, although incompatible with the Heisenberg indetermination relation, the semiclassical approach accounts for the Pauli
exclusion principle which represents the most relevant feature exhibited by
Fermi gases.
In Eq. (\ref{NF}) we have introduced the single particle  density of states 
 \begin{equation}
g(E) = {\frac{1 }{(2\pi \hbar)^3}}\int d{\bf r} d{\bf p} \delta(E-\epsilon^{sp}({\bf r},{\bf p}))
\label{gE}
\end{equation}
where $\epsilon^{sp}({\bf r},{\bf p})=p^2/2m+V_{ho}({\bf r})$ is the classical energy. The density of states depends on the dimensionality as well as on the actual form of the confining potential.
For the harmonic trapping potential (\ref{Vho}) we find 
\begin{equation}
g(E) = {1 \over 2(\hbar \omega_{ho})^3} E^2
\label{gEho}
\end{equation}
where $\omega_{ho}=(\omega_x\omega_y\omega_z)^{1/3}$ is the geometrical average of the three trapping  frequencies.
The resulting energy dependence differs from the one of the uniform 3D gas where the density of states takes the value 
$g(E) = \sqrt{E}V(m^{3/2} / (\sqrt2 \hbar^3 \pi^2))$. 
 The difference has its physical origin in the suppression of states in phase space due to the spatial confinement produced by the oscillator potential.

In terms of the density of states one can easily calculate the  relevant thermodynamic functions. For example
the energy of the gas is given by the expression
\begin{equation}
E(T)= \int_0^\infty d\epsilon {\epsilon g(\epsilon) \over e^{\beta(\epsilon -\mu)}+1} \; .
\label{ET} 
\end{equation}
 At $T=0$ the chemical potential defines  the Fermi energy ($\mu(T=0)= E_F$) and the normalization condition yields the result
\begin{equation}
E_F = k_BT_F=(6N_\sigma)^{1/3}\hbar\omega_{ho}
\label{EF}
\end{equation}
which fixes an important energy (and tempertaure) scale in the problem, analog to  expression
$E_F = (\hbar^2/2m)(6\pi^2 n_\sigma)^{2/3}$
holding for uniform matter where $n_\sigma$ is the density of a single spin component. 
 
It is worth noticing that the Fermi energy (\ref{EF}) has the same
dependence on the number of atoms and on the oscillator frequency $%
\omega_{ho}$ as the critical temperature for Bose-Einstein condensation, given by the well know formula
$k_BT_c= 0.94 \hbar \omega_{ho}N^{1/3}$. This is not a surprise since in a gas the effects of quantum
degeneracy become important when the thermal wave length $%
(2\pi\hbar^2/(mk_BT)^{1/2}$ is of the order of the average distance $n^{-1/3}(0)$
between atoms where $n(0)$ is the density in the center of the trap.
Using a classical Gaussian distribution to provide an estimate of the density  of the gas
one finds $ n^{-1/3}(0)
\sim N^{-1/3} (k_BT/m\omega_{ho}^2)^{1/2}$, so that the scale of
temperatures relevant for observing quantum effects is given by $\hbar\omega_{ho}
N^{1/3}/k_B$ in both Fermi and Bose gases. In typical experiments the value of $T_F$ corresponds to microkelvin or fractions of microkelvins.  It is however worth noticing that, differently from $T_{BEC}$, the Fermi energy does not define the critical temperature of a phase transition, but just a crossover characterizing the onset of quantum degeneracy phenomena. The occurrence of a phase transition in a Fermi gas can be only the result of interaction effects.

An important quantity to calculate is also the release energy $E_{rel}$ given by the energy of the gas after switching off the confining potential. A consequence of the equipartition theorem applied to the
ideal gas trapped by the harmonic potential is that the release energy
is always equal to $E_{rel}=E(T)/2$ where $E(T)$ is the total energy.  At $T=0$ one has $E(0)= (3/4)N_\sigma E_F$.
In Fig.1 we show how the release energy (\ref{ET}) varies
as a function of the temperature. The comparison with the prediction of the
classical gas (dashed line) and of the ideal Bose gas 
reveals explicitly the effects of quantum statistics on this measurable
quantity. These  effects were clearly demonstrated in the experiment of 
\cite{jin1,jinN} which are reported in Fig.2.

The Fermi energy (\ref{EF}) can be used to define typical length and
momentum scales characterizing the Fermi distribution in coordinate and
momentum space respectively:
\begin{equation}
E_{F}={\frac{1}{2}}m\omega _x^{2}R_x^{2}={\frac{1}{2}}m\omega _y^{2}R_y^{2}={\frac{1}{2}}m\omega
_{z}^{2}R_z^{2}={\frac{p_{F}^{2}}{2m}}  \label{EFkF}
\end{equation}
where $R_x$, $R_y$ and $R_z$ are the  widths of the density
distribution at zero temperature which can be directly calculated
integrating the $T=0$ distribution function  $f({\bf r},{\bf p})=\theta(\epsilon({\bf r},{\bf p})-E_F)$ in momentum space:
\begin{equation}
n_{0}({\bf r})={\frac{8}{\pi ^{2}}}{\frac{N_\sigma}{R_xR_yR_z}}\left( 1-
{\frac{x^{2}}{R_x^{2}}}-{\frac{y^{2}}{R_y^{2}}}-{\frac{z^{2}}{R_z^{2}}}\right) ^{3/2}\; ,
\label{n0r}
\end{equation}
while the Fermi momentum $p_{F}$ fixes the width of the corresponding
momentum distribution
\begin{equation}
n_{0}({\bf p})={\frac{8}{\pi ^{2}}}{\frac{N_\sigma}{p_{F}^{3}}}\left( 1-{\frac{p^{2}%
}{p_{F}^{2}}}\right) ^{3/2}  \label{n0k} \; ,
\end{equation}
obtained by integrating the $T=0$  distribution function  in coordinate space. Equations (\ref{n0r}) and (\ref{n0k}) hold for positive values of their
arguments and are often referred to as Thomas-Fermi distributions. Equation (\ref{n0k}) is the analogue of the most familiar
momentum distribution $3N_\sigma/(4\pi p_{F}^{3})\Theta (1-p^{2}/p_{F}^{2})$
characterizing the uniform Fermi gas. The smoothing of $n_{0}(p)$ with respect
to the uniform case is the consequence of the harmonic trapping.  Notice that
the value of $p_{F}$ defined above coincides with the Fermi momentum 
\begin{equation}
p_{F}=\hbar (6\pi ^{2}n_\sigma)^{1/3}
\label{pF}
\end{equation}
of a uniform gas evaluated in the center of trap. Using Eqs.(\ref{EF}) and (\ref{EFkF}) one easiliy finds the useful expressions 
\begin{equation}
R_i = a_{ho}\left(48 N_\sigma\right)^{1/6}{\omega_{ho} \over \omega_i}
\label{RFG}
\end{equation} 
and 
\begin{equation}
p_F= {\hbar \over a_{ho}}1.91 N_\sigma^{1/6}
\label{pFN}
\end{equation}
for the Thomas-Fermi radii and for the Fermi momentum, respectively, 
where $a_{ho}=\sqrt{\hbar /m \omega_{ho}}$ is the average oscillator length. 
It is worth comparing equations (\ref{n0r}-\ref{RFG})
 with the analogous results  holding for a trapped
Bose-Einstein condensed gas in the Thomas-Fermi limit \cite{RMP}.  For example the density distribution of the BEC configuration is given by the expression
\begin{equation}
n_0= {15\over 8\pi} {N \over R_xR_yR_z}\left( 1-
{x^2 \over R_x^2}-{y^2 \over R_y^2}-{z^2 \over R_z^2}\right)
\label{nBEC}
\end{equation}
with the Thomas-Fermi radii given by $R_i=a_{ho}(15N a/a_{ho})^{1/5}\omega_{ho}/\omega_i$ where  $a$ is the s-wave scattering length characterizing the interaction between  bosons.  The density profiles (\ref{n0r}) and (\ref{nBEC}) do not look dramatically different. In
both cases the radius of the atomic cloud increases with the number of atoms although the
explicit dependence is slightly different ($N^{1/5}$ for bosons and $N^{1/6}$
for fermions). 
 Notice, however, that the form of the density profiles has a
deeply different physical origin in the two cases. For bosons it is fixed by
the two-body interaction, while in the Fermi case it is determined by quantum pressure effects.

In momentum space the Bose and Fermi distributions instead differ in a
profound way. First, as a consequence of the semiclassical picture, the momentum distribution of the Fermi gas is isotropic even if the trapping potential 
is deformed, 
differently from what happens in the BEC case.  Second, the momentum width of a trapped Bose-Einstein condensed gas
scales like $1/R$ and hence decreases by increasing $N$ while, according to
eqs.(\ref{EFkF}-(\ref{pFN}) the momentum width of a trapped Fermi gas scales like $R$ and hence increases with the number of atoms. The different behavior reflects the
fact that the Heisenberg uncertainty inequality is close to an identity for
a Bose-Einstein condensate, while in a Fermi gas one has $\Delta P\Delta
R\sim E_{F}/\omega_{ho}\gg \hbar $.

Despite its simplicity    the ideal Fermi gas can exhibit non trivial features. This is the case  if one consider the combined presence of  harmonic plus periodic potentials which  gives rise to Bloch oscillations \cite{blochFI} and to insulating phenomena \cite{pezze} of relatively easy experimental access. Furthermore the effects of Fermi statistics of the ideal Fermi gas show up in a non trivial way in an anti-bunching behaviour exhibited by the pair correlation function which has been  the object of recent measurements \cite{Aspect,Bloch}.

\section{Role of interactions: the BEC-BCS crossover}

The ideal gas model presented in Sect. 2 provides a good description of a cold spin polarized 
Fermi gas. In this case interactions are in fact strongly inhibited by the Pauli exclusion principle. 
When  atoms occupy different spin states interactions instead deeply affect the solution of the 
many-body problem. This is particularly true at very low temperature where they give rise to pairing effects bringing the system into the 
superfluid  phase. 

Let us consider the simplest case of a two-component system occupying two different spin states 
hereafter called, for simplicity, spin-up ($\sigma=\uparrow$) and spin-down ($\sigma=\downarrow$). 
We will consider the grand canonical many-body Hamiltonian 
\begin{eqnarray}
H &=& \sum_\sigma  \int d{\bf r}\;\hat{\Psi}_\sigma^\dagger({\bf r})\left(-{\hbar^2\nabla^2 
\over 2 m}+V_{ext}({\bf r})-\mu \right)\hat{\Psi}_\sigma({\bf r}) \nonumber \\
&+&  \int d{\bf r} d{\bf r}^\prime V({\bf r}-{\bf r}^\prime)\hat{\Psi}_\uparrow^\dagger({\bf r})
\hat{\Psi}_\downarrow^\dagger({\bf r}^\prime)\hat{\Psi}_\downarrow({\bf r}^\prime)
\hat{\Psi}_\uparrow({\bf r}) \;,
\label{Hspin}
\end{eqnarray}
written in second quantization where $\mu$ is the chemical potential and the field operators obey the fermionic anticommutation relations 
$\{\hat{\Psi}_\sigma({\bf r}),\hat{\Psi}_{\sigma^\prime}^\dagger({\bf r}^\prime)\}=
\delta_{\sigma,\sigma^\prime}\delta({\bf r}-{\bf r}^\prime)$. The external potential $V_{ext}$ 
and the two-body potential $V$ account, respectively, for the external confinement and for the 
interaction between atoms of different spin. The number of atoms, fixed by the normalization condition 
$\int d{\bf r} <|\hat{\Psi}_\sigma({\bf r})|^2>=N_{\sigma}$, can in general be different for the two spin species. In this section we will consider the unpolarized case $N_\uparrow=N_\downarrow=N/2$ in the absence of external trapping ($V_{ext}=0$). The inclusion of  harmonic trapping will be discussed in the next section.

We are interested in dilute gases where the range of the interatomic potential is much 
smaller than the interparticle distance. Furthermore, we assume that the temperature of the system is 
sufficiently small so that only collisions in the $s$-wave channel are important. Under these conditions 
one expects that interaction effects will be governed by a single parameter: the $s$-wave scattering length 
$a$. In this regard one should recall that the gaseous phase corresponds to a metastable solution of the 
many-body problem, the true ground state being  in general a crystal configuration where the 
microscopic details of correlations are important (to simplify the notation the metastable solution will be  often also called the ground 
state). The above considerations explain why in 
Eq.~(\ref{Hspin}) we have ignored the interaction between atoms occupying the same spin state which, 
in most cases, is expected to give rise only to minor corrections, due to the quenching effect produced by 
the Pauli principle. Of course this picture can change considerably in the presence of $p$-wave resonances.

In order to investigate in a more tractable way the effects of quantum correlations and better understand 
the role played by the scattering length, it is convenient to replace the microscopic potential $V$ with 
an effective potential $V_{eff}$. Different types of effective potentials can be considered. In many 
applications one introduces the regularized zero-range pseudo-potential \cite{pseudopot}.
\begin{equation}
V_{eff}({\bf r})= g\delta({\bf r}){\partial\over\partial r}r \;,
\label{regpot}
\end{equation}
where the coupling constant $g$ is related to the $s$-wave scattering length $a$ characterizing the interaction between two atoms of different spin by the relationship 
$g=4\pi \hbar ^{2}a/m$. The regularization accounted for by the term $(\partial / \partial r)r$ permits 
to cure the ultraviolet divergencies in the solution of the many-body Schr\"odinger equation that 
arise from the vanishing range of the pseudo-potential.
In general, the regularization is crucial to solve the many-body problem beyond lowest order perturbation 
theory, as happens, for example, in the BCS theory of superfluidity.

In this work we will often discuss the predictions of quantum 
Monte Carlo simulations. In these numerical approaches the use of the effective potential (\ref{regpot}) is hard 
to implement and one must resort to a different effective interatomic potential. A  convenient choice is
the attractive square-well potential defined as follows: $V_{eff}(r)=-V_0$ for $r<R_0$ and $V_{eff}(r)=0$ 
otherwise [other choices for $V_{eff}$ have also been considered in the literature \cite{carlson}. The interaction 
range $R_0$ must be taken much smaller than the inverse Fermi wave vector, $k_FR_0\ll 1$, in order to ensure that 
all the many-body properties of the system will not depend on its value. The depth $V_0$ of the potential is instead 
varied so as to reproduce the actual value of the scattering length according to the relation 
$a=R_0[1-\tan(K_0R_0)/(K_0R_0)]$, where $K_0^2=mV_0/\hbar^2$. Notice that $a$ becomes infinite for every new 
two-body bound state entering the well.  

The above approaches permit to describe the many-body features uniquely in terms of the scattering length $a$. 
These schemes are adequate if the scatterng length is the only relevant interaction parameter 
and the two-body scattering amplitude, given by the expansion
\begin{equation}
f=-{1\over a^{-1}+ik+R^*k^2}
\label{amplitude}
\end{equation}
 can be safely evaluated by keeping only the first two terms in the denominator and neglecting the term containing the effective range $R^*$ of the interatomic potential. When  $R^*$ becomes of the order of 
the inverse Fermi wave vector, as happens in the case of narrow Feshbach resonances, more complex effective 
potentials should be introduced in the solution of the many-body problem \cite{bruun}.

In this section we will focus the discussion on the case of uniform systems ($V_{ext}=0$) where exact solutions of the many body problem  are available in some important limiting cases. 
A first example is the dilute repulsive gas. In this case interactions can be treated through the effective potential (\ref{regpot}) 
with a positive scattering length $a$. Standard perturbation theory can be applied with the small parameter 
$k_Fa\ll 1$ expressing the diluteness condition for the gas. The expansion of the energy per particle up to terms 
quadratic in the dimensionless parameter $k_Fa$ then  yields, at $T=0$, the result \cite{HYL}
\begin{equation}
\frac{E}{N}= \frac{3}{5}E_F \left(1+\frac{10}{9\pi}k_Fa +\frac{4(11-2\log2)}{21\pi^2}(k_Fa)^2 ...\right) \;
\label{enexpansion}
\end{equation}
where $E_F$ is the Fermi energy. The above result is universal as it holds for any interatomic potential with a 
sufficiently small effective range $R^\star$ such that $nR^{\star 3}\ll 1$. Higher order terms in the expansion
(\ref{enexpansion}) will not depend only on the scattering length $a$, but will contain an explicit dependence on 
the details of the potential. In the case of purely repulsive potentials, such as the hard-sphere model, the 
expansion in Eq.~(\ref{enexpansion}) corresponds to the energy of the ``true'' ground state of the system. For more 
realistic potentials with an attractive tail, the above result describes instead the metastable gas-like state of 
repulsive atoms. This distinction is particularly important in the presence of bound states in the two-body problem,
as more stable many-body configurations could consist of a gas of dimers (see discussion below) with the same positive value of $a$. It is finally worth 
noticing that the repulsive gas described by the equation of state (\ref{enexpansion}) does not exhibit superfluidity.

A second important case that can be solved exactly is the dilute Fermi gas interacting with negative  scattering length ($k_F|a|\ll 1$).   In this limit, hereafter called the BCS limit, the many-body problem can be solved  both at $T=0$ and at finite temperature and corresponds to the most famous BCS picture first introduced to describe the phenomenon of superconductivity. There are many variants of the BCS theory available in the literature.  We will report here the predictions of the {\it complete} BCS scheme which accounts for non trivial renormalizations of the physical quantities associated with quantum and  thermal fluctuations.  The many body solution proceeds through a proper diagonalization of the Hamiltonian (\ref{Hspin},\ref{regpot})  by applying the  Bogoliubov transformation to the Fermi field operators. This approach is non perturbative and the ground state differs in a profound way from the uncorrelated wave function of the ideal Fermi gas, being characterized by peculiar correlations associated with long range order.  The ground state energy, expanded in terms of the small parameter $k_Fa$, takes the same form (\ref{enexpansion}) holding for the repulsive gas,   pairing effects being responsible only for higher order exponential corrections. Of course in this case the first correction, linear in $a$, gives a negative contribution to the ground state energy.   BCS theory  predicts a phase transition associated with the occurrence of long range order. The corresponding  critical temperature is given by the result  \cite{melik},
 \begin{equation}
T_c=0.28 e^{2\pi/k_Fa}T_F
\label{TCTFBCS}
\end{equation}
showing that the critical temperature  becomes exponentially small as one decreases the value of $k_F|a|$, making the observability of superfluid phenomena a difficult task in dilute samples. Actually, in the experimentally relevant case case of  harmonically trapped configurations the predicted value for the critical temperature easily becomes even smaller than the  oscillator temperature $\hbar \omega_{ho}/k_B=T_F/(6N_\sigma)^{1/3}$. For example,  choosing $k_Fa=-0.2$, one finds
$T_c\sim 10^{-4}T_F$ a value significantly smaller than  the oscillator temperature for realistic values of $N_\sigma$. 

Thanks to the Fesbach resonances exhibited by several atomic species it is now possible to change the value of the scattering length in a highly controlled way by simply tuning the external magnetic field. For example, starting from a negative and small value of $a$ it is possible to increase the size of the scattering length, reach
 the resonance where $a$ diverges and explore  the other  side of the resonance  where the scattering length becomes  positive and eventually small. One would naively expect to reach in this way the regime of the dilute repulsive gas discussed above. This is not the case because in the presence of the Feshbach resonance the positive value of the scattering length is associated with the emergence of a bound state  and the formation of dimers.  The size of the dimers is of  the order of the scattering length $a$ and their binding energy is $\sim \hbar^2/ma^2$. These dimers have a bosonic nature, being composed of two fermions and if the gas is sufficiently dilute they consequently  give  rise to Bose-Einstein condensation, as experimentally proven in \cite{BECM}. The size  of the dimers cannot be however too small otherwise collisions between dimers give rise to transitions  to deeper molecular states \cite{Petrov}.  
The   behaviour  of the dilute gas of dimers ($k_Fa\ll 1$), hereafter called the BEC limit,  is properly described by the theory of Bose-Einsten condensed gases  available in both uniform and harmonically trapped configurations \cite{RMP,BookDK,BookTN}. In particular  we can immediately evaluate the critical temperature $T_c$. In the uniform case this is given by the text-book relationship $T_c= (2\pi\hbar^2/M)(n_\sigma/g_{3/2}(1))^{2/3}$ where $n_\sigma$ is  the density of dimers (equal to the density of each spin species), $M=2m$ is the mass of the dimers and $g_{3/2}(1)= 2.612$. In terms of the Fermi temperature of the uniform gas one can write 
\begin{equation}
T_c=0.22T_F
\label{TCTFunif}
\end{equation}
showing that the critical temperature for the onset of superfluidity, associated with the Bose-Einstein condensation of dimers, takes place at temperatures of the order of the Fermi temperature, i.e. at temperatures much higher than the exponentially small value (\ref{TCTFBCS}) characterizing the BCS regime. For this reason one often speaks of high $T_c$ superfluidity.
The inclusion of interactions between dimers, fixed by the molecule-molecule scattering lenght $a_M$ according to the relationship $a_M=0.6a$ \cite{Petrov}, is also straightforward and is provided, at $T=0$, by  the  Gross-Pitaevskii theory.  

The gas of dimers  and the gas of repulsive fermions discussed above represent two different branches of the many-body problem, both corresponding to positive values of the scattering length. The physical implementation of the repulsive gas configuration can be achieved by switching on adiabatically the value of the scattering length starting from the value $a=0$ (see  \cite{bourdel}). Viceversa the gas of dimers is naturally implemented by crossing the Feshbach resonance starting from negative values of $a$ (experimentally it is also realized by cooling down a gas at fixed and positive  value of the scattering length). The gas of dimers exploits the attractive nature of the force which is crucial in order to ensure the binding of the fermionic pairs. Furthermore, differently from the repulsive Fermi gas, at low temperature the gas of dimers exhibits Bose-Einstein condensation and superfluidity.

A more difficult problem concerns the behavior of  the many body system when the scattering length becomes larger than the interparticle distance. This corresponds to the challenging situation of a dilute (in the sense that the range of the force is much smaller than the interparticle distance) but strongly interacting system. Will the system be stable or will it collapse as happens for bosons interacting with large scattering lengths? 
At present an exact solution of the  many-body problem is not available for $k_F|a| >1$ and one has to make use of approximate schemes or numerical simulations. Our present understanding, based on both theoretical estimates and experimental results, is that the system  remains stable even in the so called unitary limit, 
corresponding to $k_F|a| \gg 1$. Furthermore in this limit  new interesting  features are expected to take place. In fact  the thermodynamic quanties should no longer depend on the actual value of the scattering length,   the only relevant lenght scales of the problem being   the inverse of the Fermi wave vector and  the thermal wave length \cite{ho}. An important consequence is that, at $T=0$,   the chemical potential can be parametrized in the simple way:
\begin{equation}
\mu = (1+\beta)E_F= (1+\beta) (3\pi^2)^{2/3}{\hbar^2\over 2m}n^{2/3}
\label{muunitarity}
\end{equation}
where $\beta$ is  a universal dimensionless parameter \cite{Bertsch,Heiselberg} and $n=n_\uparrow+n_\downarrow$ is the total density. The relationship (\ref{muunitarity})   fixes the density dependence of the equation of state with non trivial consequence on the behaviour of the density profiles and of the collective frequencies of harmonically trapped  superfluids (see the next sections).
  
In general we have  no reason to doubt that, at $T=0$, the system will be superfluid  for all values of $k_Fa$, i.e. along the so called BEC-BCS crossover provided by the Feshbach resonance. This will be actually the basic  point  guiding  our discussions in the following sections. The above results  make also plausible to assume that for broad resonances, corresponding to $k_FR^*\ll 1$,   all the relevant properties of the system can be described in terms of the dimensionless combination $k_Fa$. This introduces a remarkable simplification in  the theoretical description of this non trivial many-body problem.

While the exact solution of the many-body problem along   
along the BCS-BEC crossover is not available, a useful approximation is provided by the  BCS mean-field theory. This approach was first introduced to investigate the crossover by Eagles, Leggett and others   \cite{Leggett} with the main 
motivation to explore the properties of superconductivity and superfluidity beyond the weak coupling limit $k_F|a|\ll 1$. 
The main merit of this approach is that it provides a comprehensive, although approximate, description of the 
equation of state along the whole crossover, including the unitary limit $1/k_Fa \to 0$ and the BEC regime of small and positive 
$a$. 

The idea of BCS mean field theory is based on the reduction of the microscopic Hamiltonian to a mean field form where only the anomalous average $<\hat{\Psi}\hat{\Psi}>$ is taken into account, while the normal average $<\hat{\Psi}^\dagger\hat{\Psi}>$ is neglected in order to ensure convergency to the resulting equations along the whole crossover.The anomalous average characterizes the occurrence of long range order in the off-diagonal 2-body density matrix and the non vanishing value of  the pairing function
\begin{equation}
F({\bf r},{\bf s}) = <\hat{\Psi}_\uparrow({\bf r}+{{\bf s}\over 2})\hat{\Psi}_\downarrow({\bf r}-{{\bf s}\over 2})>
\label{orderparameter}
\end{equation}
The diagonalization of the resulting Hamiltonian  is obtained with the help of the Bogoliubov transformation which transforms particles into quasi-particles.  This procedure, applied to the regularized potential (\ref{regpot}), yields the non trivial  equation
\begin{equation}
\frac{m}{4\pi\hbar^2 a}= \int \frac{d{\bf k}}{(2\pi)^3}\left( \frac{m}{\hbar^2k^2} - 
\frac{1}{2E_k}\right) \;,  
\label{BCS1}
\end{equation}
where 
\begin{equation}
E_k=\sqrt{\Delta ^{2}+\eta _{k}^{2}} \;.
\label{elemexcit}
\end{equation}
is the energy of the elementary excitations  fixed by the order parameter $\Delta$ and 
\begin{equation}
\eta = {\hbar^2 k^2 \over 2m} -\mu
\label{etamu}
\end{equation}
is the energy of a free particle calculated with respect to the chemical potential. The order parameter is directly related to the pairing function (\ref{orderparameter}) through the relation
\begin{equation}
\Delta = -\int d{\bf s}\; V_{eff}({\bf s}) <\hat{\Psi}_\uparrow({\bf r}+{{\bf s}\over 2})\hat{\Psi}_\downarrow({\bf r}-{{\bf s}\over 2})>
\label{DBCS}
\end{equation} 
where $V_{eff}$ is the pseudopotential (\ref{regpot}). For positive values of the chemical potential the order parameter $\Delta$ coincides with the energy gap of the particle excitation spectrum (\ref{elemexcit}). For negative values of $\mu$ the gap is instead given by $\sqrt{\Delta^2 + \mu^2}$. The gap has been measured in trapped gases \cite{grimmgap}  along the BEC-BCS crossover through  radio-frequency transitions. 

Equation (\ref{BCS1}) provides an important 
relationship between $\Delta$ and the chemical potential $\mu$ entering the single-particle energy $\eta_k$. 
A second relation is given by the normalization condition which is preserved by the Bogoliubov transformation and takes the form: 
\begin{equation}
n=\int \frac{d{\bf k}}{(2\pi)^3}\; \left(1-\frac{\eta_k}{E_k}\right) \;.
\label{BCS2}
\end{equation}

The two equations should be solved in a consistent way. It is remarkable that they admit a solution for any value of $k_Fa$. The resulting prediction for the equation of state $\mu(n)$ given by the BCS-mean field theory is reported in Fig.3 where it is compared with the {\it ab initio} calculation of \cite{stefano}, based on a Monte Carlo simulation, and the selfconsistent calculation of \cite{strinatiT} based on a diagrammatic expansion.
The figure reveals a general qualitative agreement  between the two approaches. In particular in the BEC limit the mean field theory accounts for the existence of bound molecules of energy $\hbar^2/ma^2$. The BCS mean field theory however misses some important features which are worth being mentioned:

i) in the BEC limit the interaction between molecules is known to be governed by the value $a_M=0.6 a$ of the molecule-molecule scattering length \cite{Petrov}. BCS theory instead provides the wrong value $a_M=2a$

ii) In the BCS limit the mean field BCS theory does not account for the leading corrections to the ground state energy
(see eq.(\ref{enexpansion})). This is due to the fact that the theory ignores the Hartree terms $<\hat{\Psi}^\dagger\hat{\Psi}>$ in the calculation of the energy.

iii) the value of the order parameter $\Delta$ in the BCS regime misses the Gorkov-Melik Barkhudarov correction which reduces the proportionality coefficient between $\Delta$ and the Fermi energy  by a significant factor ($\sim 0.45$).

iv) At unitarity the BCS theory predicts the value $\beta=-0.42$ to be compared with the value $\beta=-0.58$ given by the numerical Monte Carlo simulation.

\section{Equilibrium properties of a trapped gas}

Let us first  point out that in the deep BEC regime, where the interacting Fermi gas behaves like a gas of  weakly interacting dimers,  systematic  information is available from our advanced knowledge of the physics of dilute Bose gases in harmonic traps \cite{RMP}. Although the BEC regime is not easily achieved in present experiments with ultracold  Fermi gases, the corresponding predictions nevertheless provide a useful reference for important physical quantities. For example the critical temperature for BEC in a harmonic trap is given by the well known expression $k_BT_{BEC}=0.94 \hbar \omega_{ho}N_\sigma^{1/3}$ which, combined with the corresponding expression (\ref{EF}) for the Fermi energy, provides the useful relationship
\begin{equation}
T_{BEC} =0.52 T_F
\label{TBECtrap}
\end{equation}
Result (\ref{TBECtrap}) reveals that, once expressed  in units of the Fermi temperature, the critical temperature in the BEC regime is higher than the corresponding value (\ref{TCTFunif}) holding in uniform matter. In this sense superfluidity is "favoured" by the presence of the harmonic trap.
The critical temperature in trapped Fermi gases can be estimated also in other regimes, using the results of uniform matter within local density approximation. First estimates of $T_c$ in the BCS regime were given in \cite{stoof}.

In the following we will show how the results of the previous section can be usefully employed to describe the equilibrium properties of a trapped gas along the BEC-BCS crossover at zero temperature. To this purpose we will make use of the local density approximation (LDA). This approximation assumes that, locally, the system behaves like a uniform body so that the energy of the trapped system can be written in the integral form  
\begin{equation}
E =  \int d{\bf r} \left[\epsilon(n({\bf r})+V_{ho}({\bf r})n({\bf r}) \right]
\label{Etot}
\end{equation}
given by the sum of the internal, also called release,  energy
\begin{equation}
E_{rel}=\int d{\bf r}\epsilon(n({\bf r}))
\label{Erel}
\end{equation}
and  of the oscillator energy 
\begin{equation}
E_{ho}=\int d{\bf r}V_{ho}({\bf r})n({\bf r})
\label{Epot}
\end{equation}
provided by the trapping potential (\ref{Vho}).
 In  eqs.(\ref{Etot}-\ref{Epot}) $n({\bf r})=n_\uparrow({\bf r})+n_\downarrow({\bf r})$  is the  total  density profile determined by the variational relation $\delta (E-\mu_0N)/ \delta n({\bf r})=0$ which  yields the most relevant LDA equation
 \begin{equation}
\mu_0=  \mu(n({\bf r})) + V_{ho}({\bf r})
\label{LDA}
\end{equation}
where 
\begin{equation}
\mu(n)={\partial \over \partial n}\epsilon(n)
\label{muE}
\end{equation}
is the chemical potential of uniform matter and $\mu_0$ is the chemical potential of the trapped gas, fixed by the normalization condition $\int d{\bf r}n({\bf r})=N$. 
Equation (\ref{LDA}) provides an implicit equation for the density profile $n({\bf r})$ holding at equilibrium.  

The applicability of LDA in a Fermi gas is justified if the external potential is sufficiently smooth (which is in general guaranteed by the harmonic choice) and  if the relevant energies are much larger than the typical oscillator energies  providing the quantization of the single particle levels ($\mu_0 \gg \hbar \omega_0$). In the absence of interactions the equation of state is given by the ideal Fermi gas expression 
\begin{equation}
\mu(n) = (3\pi^2)^{2/3}{\hbar^2\over 2m}n^{2/3}
\label{munF}
\end{equation}
yielding the result 
\begin{equation}
n({\bf r})={1\over 3\pi^2}({2m \over \hbar^2})^{3/2}(\mu_0-V_{ho}({\bf r})^{3/2}
\label{nmuF}
\end{equation} 
for the equilibrium density profile. For harmonic trapping this corresponds to the $3/2$-th power of an inverted parabola (see eq.(\ref{n0r})). The chemical potential
$\mu_0$  takes the value
\begin{equation}
\mu_0 = \hbar\omega_{ho}(6N_\sigma)^{1/3} \equiv E_F
\label{mu0F}
\end{equation}
and, as expected, coincides  with the Fermi energy (\ref{EF}) already introduced for the trapped ideal gas.

Interactions  modify the shape and the size  of the density profiles. The effects are accounted for by eq.(\ref{LDA}) once  the  equation of state $\mu(n)$ is known.  A simple result is obtained at unitarity where the equation of state (see eq.(\ref{muunitarity})) has the same density dependence (\ref{munF}) as for the ideal gas, apart from a dimensionelss renormalzation factor. By dividing (\ref{LDA}) by $(1+\beta)$ one then finds that  the results at unitarity are simply obtained from the ones of the ideal Fermi gas by  a simple rescaling of  the trapping frequencies and of the chemical potential.   In particular the density profile at unitarity takes the same form  
\begin{equation}
n_{0}({\bf r})={\frac{8}{\pi ^{2}}}{\frac{N}{R_xR_yR_z}}\left( 1-
{\frac{x^2}{R_x^2}}-{\frac{y^2}{R_y^2}}-{\frac{z^2}{R_z^2}}\right)^{3/2}
\label{nunitarity}
\end{equation}
as in the ideal gas, with the Thomas-Fermi radii given by the rescaled law
\begin{equation}
R_i = (1+\beta)^{1/4}R_i^0 
\label{Runitarity}
\end{equation}
where $R_i^0$   are the Thomas-Fermi radii of the ideal gas (see eq. (\ref{RFG})). 

Another important  case is the BEC limit where one treats the interaction between dimers using the mean field equation of state  $\mu=g_Mn/4$  with the coupling constant $g_M=4\pi\hbar^2a_M/2m$  fixed by the molecule-molecule scattering length. According to the well known results holding for weakly interacting bosons one finds 
\begin{equation}
R_i = a_{ho}({15\over 2} N{a_{M}\over a_{ho}})^{1/5}{\omega_{ho}\over \omega_i}
\label{RBEC}
\end{equation}

In Fig.4 we show the experimental results \cite{grimmdensity} for the {\it in situ }  density profiles in a harmonically trapped Fermi gas at extremely low temperature. The results actually correspond to the axial column density, i.e. to the integral 
$\int dx dy  n({\bf r})$ which is the quantity  measured in the experiment of \cite{grimmdensity}. Very good agreement between eperiment and theory  is found at unitarity where the column  integral of (\ref{nunitarity})  is easily calculated and yields the result:
\begin{equation}
n_{column}(z)= {16\over 5\pi}{N\over Z}(1 -{z^2 \over Z^2})^{5/2}
\label{column}
\end{equation}
with $Z\equiv R_z$ given by eq.(\ref{Runitarity})
The best fit to the experimental curve yields the value $\beta=-0.68$ revealing explicitly  the attractive role of interactions at unitarity. In Fig.5 we show the experimental results  for the axial radius of the trapped cloud along the crossover obtained in the same experiment. 
The results on the BEC side  are reasonably consistent with the prediction (\ref{RBEC}) (continuous line) once one uses the value $a_M=0.6 a$, provided by the exact solution of the dimer-dimer scattering problem \cite{Petrov}. For a systematic comparison between experimental and theoretical results of the density profiles see \cite{strinati}.

In addition to the {\it in situ} density profiles a valuable source of information comes from the measurement of the release energy, i.e. the energy of the system measured after switching off the confining trap. The release energy is defined by the sum 
\begin{equation}
E_{rel}=E_{kin}+E_{2-body}
\label{release}
\end{equation}
of the kinetic and the 2-body interaction terms and, in the local density approximation, is simply given by eq.(\ref{Erel}), i.e. by the first term of the total energy (\ref{Etot}).
The integral (\ref{Erel}) should be in general carried out numerically. There are however  important cases where  analytical results are available.  
For example, at unitarity,  one finds the useful result
\begin{equation}
E_{rel}= {3 \over 8}(1+\beta)^{1/2}NE_F
\label{releaseunitarity}
\end{equation}
where $E_F=\hbar\omega_{ho}(3N)^{1/3}$ is the Fermi energy (\ref{mu0F}) of the ideal trapped gas. 

From the measurement of the  release energy T. Bourdel et al.  \cite{BECM} were able to  extract the value $-0.64$ for the parameter $\beta$ in reasonably good agreement with the value  extracted from the {\it in situ} measurement of the profile as well as with the {\it ab initio} predictions of theory.
   
Another quantity that has been recently measured in ultracold Fermi gases is the momentum distribution \cite{jin2}. To this purpose one switches  off the scattering length, profiting of the existence of a Feshbach resonance,   just before the expansion of the gas. The images of the density profile after expansion then provide direct access to the momentum distribution. The experimental and theoretical results are shown in Fig.6 for different values of $k^0_Fa$. The theoretical curves  have been obtained by applying  the local density approximation \cite{viverit} $n(k)=(2\pi)^{-3}\int d{\bf r}n_k({\bf r})$ to the particle distribution function 
\begin{equation}
n_k\equiv <a_k^\dagger a_k>= {1\over 2}\left(1-{\eta_k\over E_K}\right)
\label{nk}
\end{equation}
calculated in uniform matter at the corresponding value of the density, within BCS mean field theory.
Eq. (\ref{nk}) reduces to the step function $\Theta(k-k_F)$ in the deep BCS limit, while it approaches the value 
\begin{equation}
n_k= {4\over 3}(k_Fa)^3{1 \over (k^2a^2 +1)^2}
\label{nkdimers}
\end{equation} 
in the opposite BEC limit where, as expected, it coincides with the momentum distribution of a particle in the bound state of the dimer.  
The figure reveals a rather satisfactory agreement between theory and experiments.

\section{Dynamics and superfluidity}

Superfluidity is one of the most challenging properties exhibited by ultracold Fermi gases, analog to the superconductor behaviour taking place in charged Fermi systems.  It shows up in peculiar transport features. Among the most noticable manifestations one  should recall the absence of viscosity, the hydrodynamic nature of macroscopic dynamics even at zero temperature, the existence of quantized vortices, the occurrence of pairing effects and the vanishing of  the spin polarizability. The last two features are typical of Fermi superfluids, while the first ones characterize also the superfluid behavior of Bose systems. The possibility of exploring these phenomena in ultracold gases provides a unique opportunity to complement our present knowledge of superfluidity in neutral Fermi systems, previously limited to liquid $^3$He. In this chapter we will discuss the hydrodynamic behaviour exhibited by superfluids and its implications on the dynamics of trapped Fermi gases. We will also discuss the dynamic behaviour   at higher energies where pair breaking effects become important.   The implications of superfluidity  on the rotational properties will be discussed in the following section. 

The macroscopic behaviour of a neutral superfluid is governed by the equations of irrotational hydrodynamics.
At zero temperature they consist  of coupled and closed equations for the density and the velocity field. In fact, due to the absence of the normal component, the superfluid density coincides with the total density and the superfluid current with the total current. The
equations take the form  
\begin{equation}
{\partial \over \partial t}n + \nabla {\bf \cdot} (n {\bf v}) =0
\label{continuity}
\end{equation}
for  the density (equation of continuity)
 and  
\begin{equation}
m{\partial \over \partial t}{\bf v} + \nabla \left({1\over 2}m{\bf v}^2 + \mu(n) + V_{ho} \right)  =0
\label{euler}
\end{equation}
for the velocity field (Euler equation) where $\mu(n)$ is the  chemical potential per particle,  fixed by the equation of state of bulk matter. At equilibrium ({\bf v}=0) the Euler equation provides the Thomas-Fermi condition (\ref{LDA}) for the ground state profile. The irrotationality of the velocity field, implied by eq.(\ref{euler}), is the consequence of the existence of the order parameter (\ref{DBCS}) whose phase $\Phi$  is related to the superfluid velocity field by the relationship
\begin{equation}
{\bf v}={\hbar \over 2m}\nabla \Phi
\label{vphase}
\end{equation}
The hydrodynamic equations (\ref{continuity}-\ref{euler}) differ from the corresponding equations holding in the collisionless regime of a non superfluid system  because of the irrotationality constraint (\ref{vphase}). 

Despite the quantum origin underlying the superfluid behaviour, the hydrodynamic equations of motion have a classical form and do not depend explicitly on the Planck constant. This peculiarity raises  the question whether the hydrodynamic behaviour of a cold Fermi gas can  be used to test the achivement of the superfluid regime. As we will see, Fermi gases above the critical temperature  can easily enter a  collisional regime where the dynamic behaviour is governed by the same   equations. In this respect it is important to stress that collsional hydrodynamics  admits the possibility of rotational components in the velocity field which are strictly absent in the superfluid.  A distinction between classical  and superfluid hydrodynamics is consequently possible only studying the rotational properties of the gas (see next section).  

The applicability of the hydrodynamic equations is restricted to the study of macroscopic phenomena, characterized by  low long wave length excitations.  In particular the wave lengths  should be larger than the so called healing length. In  
the limit of BEC dimers the healing length is proportional to $ \sqrt{\hbar^2/Mg_Mn}$ where $M=2m$ and $g_Mn$ is the molecular mean field energy. In the opposite BCS limit the healing length is instead proportional to $\hbar/\Delta$  where $\Delta$  is the pairing gap. At unitarity the healing length is of the order of the inter-particle distance. At the end of the section we will relate the healing length to the critical Landau's velocity and discuss its behavior along the BEC-BCS crossover. 
Let us finally remark that the hydrodynamic equations of superfluids have the same form both for Bose and Fermi systems,
the effects of statistics entering only  the form of the equation of state  $\mu(n)$. 

Important applications of the hydrodynamic equations concern the expansion of the gas after release of the trap and the collective oscillations.

In most experiments with ultracold atomic gases images are taken after expansion of the cloud. In the absence of interactions the expansion of a Fermi gas is asymptotically isotropic even if the gas is initially confined by an anisotropic potential. This is the consequence of the isotropy of the momentum distribution $n({\bf p})$ which, for a non interacting gas, fixes the asymptotic beavior of the density $n({\bf r},t)$ of the expanded gas  according to the law $n({\bf r},t) \to (m/t)^3n({\bf p})$ whith ${\bf r}=t{\bf p}/m$. Deviations from isotropy are consequently an important indicator of the role of interactions. 
In the experiment of \cite{ohara} et al the first clear evidence of anisotropic expansion of an interacting ultracold Fermi gas was reported, opening an important debate in the scientific community aimed to understand the  nature of these novel many-body configurations. Hydrodynamic theory  has been extensively used in the past years to analyze the expansion of  Bose Einstein condensed gases. More recently  it was proposed as a theory for the expansion of a Fermi superfluid \cite{menotti}. The hydrodynamic solutions  are obtained starting from the equilibrium configuration, corresponding to a Thomas-Fermi profile, and then solving eq.(\ref{euler}) by setting $V_{ho}=0$ for $t>0$. For an  important class  of configurations the spatial dependence  can be analytically inferred.  In fact, if the chemical potential has the power law $\mu\propto n^{\gamma}$   dependence on the density, the Thomas-Fermi equilibrium profiles have the analytic form $n_0\propto (\mu_0-V_{ho})^{1/\gamma}$ and one can easily prove that  the scaling ansatz 
\begin{equation}
n((x,y,z,t) = (b_xb_yb_z)^{-1}n_0({x\over b_x},{y\over b_y},{z\over b_z})
\label{scaling}
\end{equation}
provides the exact solution for the expansion with the scaling parameters $b_i$ obeying the simple time dependent equation:
\begin{equation}
\ddot{b}_i - {\omega^2_i \over b_i(b_xb_yb_z)^\gamma }=0 \; . 
\label{bi}
\end{equation}
Equation (\ref{bi}) generalizes the scaling equations previously introduced in the case of an interacting Bose gas ($\gamma=1$) \cite{dum}.
From the solutions of eq.(\ref{bi})  one can  easily calculate the   aspect ratio as a function of time. For an axially symmetric trap  ($\omega_x=\omega_y\equiv \omega_{\perp}; b_x=b_y \equiv b_\perp$) this is defined as the ratio between the radial and axial  radii. In terms of the scaling parameters $b_i$ it can be  written as 
\begin{equation}
{R_{\perp}(t) \over Z(t)} = {b_\perp(t)\over b_z(t)}{\omega_z \over \omega_\perp}
\label{ratio}
\end{equation}
For an ideal gas the aspect ratio tends to unity, while the hydrodynamic equations yield an asymptotic value $\neq 1$. Furthermore hydrodynamics predicts a peculiar inversion of shape  during the expansion caused by the hydrodynamic forces which are larger in the direction of larger density gradients. As a consequence  an initial cigar shaped configuration is brought into a disk profile at large times and viceversa. One can easily estimate the typical time at which the inversion of shape takes place. For a highly elongated trap ($\omega_\perp \gg \omega_z$) the axial radius is practically unchanged for short times since the relevant expansion time along the $z$-th axis is fixed by $1/\omega_z \gg 1/\omega_\perp$. Conversely the radial size increses fast, and, for $\omega_\perp t \gg 1$ one expects $R_\perp(t) \sim R_\perp(0)\omega_\perp t$. One then finds that the aspect ratio is equal to unity when $\omega_z t \sim 1$.

In Fig.7 we show the experimental images \cite{ohara} of the expanding cloud of an ultracold Fermi gas taken at different times close to unitarity. The figure clearly shows the inversion of the shape from cigar to disc  predicted by hydrodynamic theory. In Fig.8 the predictions for the aspect ratio given by eqs. (\ref{bi}-\ref{ratio}) at unitarity, where $\gamma=2/3$, are shown together with the  experimental results of \cite{ohara}.  The configuration shown in these figures corresponds to an initial aspect ratio equal to $R_\perp/Z=0.035$. The comparison strongly supports the  hydrodynamic nature of the expansion of these ultracold Fermi gases. The  experiment was  repeated at higher temperatures and found to exhibit a similar hydrodynamic behaviour even at temperatures of the order of the Fermi temperature, where the system cannot be superfluid. One then  concludes that in the normal phase the system exhibits a  collisional regime. This is especially plausible close to unitarity where the scattering length is very large. 

The collective oscillations of a superfluid gas provide a further relevant source of  information. These oscillations can be studied by considering the linearized form of the time dependent HD equations (\ref{continuity}-\ref{euler}), corresponding to small oscillations $n=n_0+\delta n\exp(-i\omega t)$ of the density with respect to the equilibrium profile $n_0$, where $\omega$ is the frequency of the oscillation. The linearized equations take the form
\begin{equation}
-\omega^2\delta n = \nabla \cdot \left[n_0\nabla ({\partial \mu \over \partial n}\delta n)\right]
\label{linearized}
\end{equation}
the velocity field being fixed by the  equation 
\begin{equation}
m{\partial {\bf v} \over \partial t} = -\nabla({\partial \mu \over \partial n}\delta n)
\label{linearizedv}
\end{equation}

Let us first consider the case of isotropic harmonic trapping ($\omega_x=\omega_y=\omega_z\equiv \omega_{ho}$). A general class of divergency free (also called surface) solutions is available in this case. They are characterized by the velocity field  ${\bf v} \propto \nabla (r^\ell Y_{\ell m})$, satisfying the condition $\nabla \cdot {\bf v}=0$ and corresponding to the behavior   $(\partial \mu/\partial n)\delta n \propto r^\ell Y_{\ell m}$ for the density variation. Using the identity $(\partial \mu/\partial n)\nabla n_0=-\nabla V_{ho}$, holding for the density profile at equilibrium, it is immediate to find that these solutions obey the equation  
\begin{equation}
\omega^2 {\partial \mu \over \partial n}\delta n = \nabla V_{ho}\cdot \nabla({\partial \mu \over \partial n}\delta n) \; .
\label{linearized}
\end{equation}
The resulting dispersion law
\begin{equation}
\omega(\ell)= \sqrt\ell \omega_{ho} .
\label{surface}
\end{equation}
is independent of the form of the equation of state, as generally expected for the surface modes driven by an external force. This result   provides a  model independent characterization of the hydrodynamic regime. The result in fact differs from the prediction   $\omega(\ell)=\ell \omega_{ho}$ of the ideal gas model, revealing the importance of interactions accounted for by the hydrodynamic description. Only in the dipole case ($\ell=1$), corresponding to the rigid oscillation of the center of mass, interactions do not affect the frequency of these modes. In addition to the surface modes an important solution predicted by the hydrodynamic equations in isotropic harmonic traps is  the $\ell=0, m=0$ breathing radial mode whose solution can be found in analytic form if the equation of state is polytropic ($\mu \propto n^{\gamma}$). In this case the velocity field has the radial form ${\bf v} \propto {\bf r}$ and the frequency is equal to 
\begin{equation}
\omega(m=0)=\sqrt{3\gamma+2}\omega_{ho} \; .
\label{m=0spherical}
\end{equation}
For $\gamma=1$ one recovers the well known BEC  result $\sqrt5\omega_{ho}$ \cite{sandro96}, while at unitarity one finds $2\omega_{ho}$. It is worth stressing that the unitary result $ \omega = 2\omega_{ho}$ is not limited to small amplitude oscillations and   keeps its validity beyond the hydrodynamic approximation \cite{castin}. 

In the case of axi-symmetric trapping ($\omega_x=\omega_y\equiv \omega_\perp\ne \omega_z$) the third component $\hbar m$ of angular momentum is still a good quantum number and one  also finds simple solutions of eq.(\ref{linearized}). The dipole modes, corresponding to the center of mass oscillation, have frequencies $\omega_\perp$ for $m=\pm1$ and $\omega_z$ for $m=0$. The oscillations where the velocity field is linear in the spatial coordinates  exhibit a richer structure. The solutions with $m=\pm 2$ and $m=\pm 1$ are surface excitations of the form ${\bf v}\propto \nabla(x\pm iy)^2$ and ${\bf v}\propto \nabla(x\pm iy)z$, with frequency given, respectively, by 
\begin{equation}
\omega(m=\pm 2)=\sqrt2\omega_\perp
\label{m=2}
\end{equation}
and
\begin{equation}
\omega(m=\pm 1)=\sqrt{\omega_\perp^2+\omega^2_z} \; ,
\label{m=1}
\end{equation}
independent of the equation of state. The dispersion (\ref{m=2}) of the radial quadruple mode has been recently tested experimentally along the crossover \cite{grimmprivate}. 

The $m=0$ solutions instead depend  on the equation of state. For  a polytropic dependence of the chemical potential  state ($\mu \propto n^\gamma$) the problem can be solved in an analytic way. The solutions are characterized by a velocity field of the form ${\bf v}\propto \nabla[a(x^2 +y^2)+bz^2]$,  resulting from the coupling  between the $\ell=2$  and $\ell=0$ modes caused by the deformation of the trap. The corresponding  frequencies are given by
\cite{cozzini} 
\begin{eqnarray}
&&\omega^2(m=0)= {1\over 2} [2(\gamma+1)\omega^2_\perp+ (\gamma+2)\omega^2_z  \pm \nonumber \\
&&\sqrt{4(\gamma+1)^2\omega^4_\perp+(\gamma+2)^2\omega^4_z+4(\gamma^2-3\gamma-2)\omega^2_\perp\omega_z^2}]
\label{m=0}
\end{eqnarray}
Equation (\ref{m=0}) reduces to the one derived in \cite{sandro96} in the interacting Bose case ($\gamma=1$), while at unitarity ($\gamma=2/3$)  it coincides with the HD prediction applied to the isoentropic oscillations of the ideal gas  \cite{Griffin,Kagan,Amoruso,bruunclark}. For elongated traps ($\omega_z\ll \omega_\perp$) the two solutions (\ref{m=0}) reduce to $\omega=\sqrt{2(\gamma+1)}\omega_\perp$ and $\omega=\sqrt{(3\gamma+2)/(\gamma+1)}\omega_z$.  
 
Let us discuss in more detail the behavior of the compression modes at unitarity where, for elongated traps, one finds the prediction $\sqrt{10/3}\omega_\perp$ and $\sqrt{12/5}\omega_z$ for the radial and axial  oscillations, respectively. 
 Experimentally both the two  modes have been investigated in ultracold Fermi gases \cite{thomasoscillations,grimmoscillations}.  In Fig. 9  we show the recent experimental results for the compressional radial mode taken from \cite{grimmoscillations}. The agreement between theory and experiment at unitarity is remarkable confirming our understanding of the dynamic behaviour in this highly correlated regime where the scattering length is much larger than the interparticle distance and the system exhibits a universal behavior. It is also worth noticing that the damping of the oscillations is smallest near unitarity. 
 
When we move from unitarity the collective oscillations  exhibits other interesting features. Theory predicts that in the BEC regime ($\gamma=1$) the frequencies of both the axial and radial modes are higher than at unitarity (equal to $2\omega_\perp$ and $\sqrt{5/2}\omega_z$ respectievly). Furthermore the first corrections with respect to the BEC prediction can be  calculated analytically, by accounting for the first correction to the BEC equation of state $\mu=gn$  produced by quantum fluctuations. This is the so called Lee-Huang-Yang (LHY) correction   first derived in the framework of Bogoliubov theory of a uniform Bose gas \cite{HYL}. The resulting shifts in the collective frequencies  can be calculated analytically by solving the hydrodynamic equations (\ref{linearized}) and treating the LHY term  in a perturbative way.  In the case of the most relevant  $m=0$ radial breathing mode in a cigar-like configuration ($\omega_z\ll \omega_\perp$) one finds the following result for the relative frequency shift \cite{levLHY,braaten}
 \begin{equation}
 {\delta \omega \over \omega}={105 \sqrt\pi\over 256}\sqrt{a^3_Mn(0)} 
 \label{PSshift}
 \end{equation}
 where $\omega=2\omega_\perp$ is the unperturbed value.
The shift is  positive reflecting the repulsive nature of the interaction between molecules As a consequence, the dispersion law, when one moves from the BEC regime towards unitarity, exhibits a typical non monotonic behavior, since it first increases, as a consequence of the LHY effect,  and eventually decreases to reach a lower value at unitarity \cite{sandro04}. In general the collective frequencies can be calculated numerically along the whole crossover by solving the hydrodynamic equations once the equation of state is known.  Fig.9 shows the predictions obtained using the equation of state of the  MC simulation \cite{combescotoscillations} (see also \cite{manini} and of   BCS mean field theory \cite{hu}. The MC equation of state accounts for the LHY effect while the mean field BCS theory misses it, providing a monotonic behavior for the compressional frequencies as one moves from the BEC regime to unitarity.  The equation of state  is in both cases consistent with the correct value $\sqrt{10/3}\omega_\perp$ at unitarity. The  accurate measurements of the radial compression mode shown in Fig.9  confirm  the prediction of the MC simulation, providing an important  test  of the equation of state and the first observation of the LHY effect.
 
 The behavior of the collective frequencies on the BCS side of the resonance exhibits different features. Theoretically one expects that when the system reaches the BCS regime the frequencies should be the same as at unitarity, the equation of state being governed by the same $2/3$ power law density dependence. However,  things behave differently in experiments. The observed behavior is not a surprise and can be qualitatively understood by noticing that when one moves towards the BCS regime the critical temperature  and the pairing gap become smaller and smaller and soon reaches values of the order of the trapping oscillator  frequencies. Under these conditions, even assuming that the system be at zero temperature, one looses superfluidity and the system is eventually expected to behave like a dilute collisionless gas  whose collective frequencies, apart from minor mean field corrections, should approach the higher values $2\omega_\perp$ and $2\omega_z$, respectively,  for the radial and axial modes. Experimentally this  transition  is  observed for the radial mode where the relevant trapping frequency is higher  and  occurs at about $k_F|a|\sim 1$. It is also associated with a strong increase of the damping of the collective oscillation.

 As already pointed out, hydrodynamic theory describes correctly only the low frequency  oscillations of macroscopic nature, corresponding to sound waves in a uniform body. When one considers higher excitation energies the dynamic response should also include the breaking of pairs into two fermionic excitations. The general picture of the excitations  produced by a density probe can then be summarized as follows (for simplicity we consider a uniform gas): at low frequency the system exhibits a gapless phononic branch whose slope is fixed by the sound velocity $c$ and hence by the compressibility of the gas according to the equation
\begin{equation}
mc^2 = n {\partial \mu \over \partial n}
\label{mc2}
\end{equation}
At high frequency one expects the emergence of a continuum of excitations starting from a given threshold frequency, above which one can break pairs. The value of the threshold frequency  depends on the value of the total momentum carried by the perturbation. A first estimate is provided by the BCS mean field theory which predicts the following result for the threshold
\begin{eqnarray}
&\hbar\omega_{th}& = 2\Delta  \; \;  {\text for} \; \;  \mu>0 \; \; {\text and} \; \;  q\le 2\sqrt{2\mu}    \nonumber \\
&\hbar\omega_{th}& =  2\sqrt{(q^2/8m-\mu)^2+\Delta^2} \; \; {\text  elsewhere}
\label{summaryexcitations}
\end{eqnarray}
The interplay between  phonon and pair breaking excitations gives rise to different scenarios  along the crossover. 
In the BCS regime the threshold  occurs at low frequencies  and the phonon branch very soon reaches the continuum of single particle excitations. The behavior is quite different in the opposite BEC regime where the gapless phonon branch extends up to high frequencies. At large momenta this branch  actually looses its phononic character and approaches the   dispersion 
$q^2/4m$ of a free molecule. In the deep BEC limit the gapless branch    coincides with the Bogoliubov spectrum of a dilute gas of bosonic molecules. At unitarity the system is expected to exhibit an intermediate behavior, the discretized branch surviving up to momenta of the order of the Fermi momentum. A detailed calculation of the excitation spectrum \cite{combescot06}, based on a proper time-dependent generalization of the mean field BCS theory, is provided in Fig. 10. 

The results for the excitation spectrum provide a useful  insight  on the superfluid behavior of the gas in terms of the Landau's criterion  according to which a system cannot give rise to energy dissipation if its velocity, with respect to a container at rest, is smaller than the Landau's critical velocity defined by the equation
\begin{equation}
v_{cr} = min_q{\hbar\omega_q\over q}
\label{criticalv}
\end{equation}
where $\hbar\omega_q$ is the energy of  an excitation carrying momentum ${\bf q}$.
According to this criterion the ideal  Fermi gas is not superfluid because of the absence of a threshold for the single particle excitations, yielding $v_{cr}=0$. The interacting Fermi gas of Fig.10 is instead superfluid in all regimes.   By inserting result (\ref{summaryexcitations})  for the threshold frequency into eq.(\ref{criticalv}) one can calculate  the critical value of $v$  due to pair breaking. The result is
\begin{equation}
v_{cr}^{sp}= \left({[\sqrt{\Delta^2+\mu^2} -\mu]\over m}\right)^{1/2}
\label{vcrsp}
 \end{equation}
 In the deep BCS limit $|a|k_F \to 0$ (corresponding to $\Delta\ll \mu$) eq. (\ref{vcrsp}) approaches the exponentially small value  $v_{cr}=\Delta/p_F$. On the BEC side the value (\ref{vcrsp}) instead becomes larger and larger and the relevant excitations giving rise to Landau's instability are no longer single particle excitations but   phonons, and   the  critical velocity coincides with the sound velocity. A simple estimate of the critical velocity along the whole crossover is then given by the  expression 
\begin{equation}
v_{cr}= Min\left(c,v_{cr}^{sp}\right)
\label{vcrcrossover}
\end{equation}
Remarkably, one sees that $v_{cr}$ has a maximum near unitarity (see Fig.11),  further confirming the robustness of superfluidity at unitarity. The large value of $v_{cr}$  should show up in a visible  reduction of dissipation  in experiments where one moves an external impurity in the medium at tunable velocities. Experiments of this type have been already performed in the case of Bose-Einstein condensed gases \cite{onofrio}.

An important physical quantity, directly related to the critical velocity, is the healing length defined as

\begin{equation}
\xi={\hbar\over m v_{cr}} \; .
\label{xi}
\end{equation} 
Apart from a trivial numerical factor it coincides with the usual definition $\hbar/\sqrt{Mg_Mn}$ of the healing length on the BEC side and with the size of Cooper pairs in the opposite BCS limit. The healing length provides the typical length scale above which the dynamic description of the system is safely described by the hydrodynamic picture. It is smallest near unitarity. 

The  knowledge of the excitation spectrum and of the corresponding matrix elements of the density operator allows one to calculate the dynamic structure factor \cite{BookTN}
\begin{equation}
S(k,\omega)= Q^{-1}\sum_{m,n}e^{-\beta E_{mn}}|<0|\delta\hat{\rho}_{k}|n>|^2\delta(\hbar\omega-\hbar\omega_{mn})
\label{Sqomega}
\end{equation}
where $\hbar {\bf k}$ and $\hbar \omega$ are the momentum and energy, respectively, transferred by the probe to the sample, $\delta\hat\rho$ is the  fluctuation of the Fourier component  $\hat\rho_k=\sum_j \exp{-i{\bf k \cdot r}_j}$  of the density operator, $\omega_{mn}=(E_m-E_n)/\hbar$ are the usual Bohr frequencies and $Q$ is the partition function. The definition of the dynamic structure factor is immediately generalized to other excitation operators   like, for example, the spin density operator. In dilute gases the dynamic structure factor can be measured with Bragg scattering experiments. 

The main features of the dynamic structure factor are best understood in uniform matter where the excitations are described in terms of their momentum. From the previous discussion on the excitation spectrum  one expects that, for sufficiently small momenta, the dynamic structure factor be characterized by a discretized peak   and by a continuum of single-particle excitations at higher energy. At higher momentum transfer the behavior will depend crucially on the regime considered along the BEC-BCS crossover.  In Fig.12 we report the $T=0$ predictions for the dynamic structure factor $S({\bf k},\omega)$ at relatively high wave vectors ($k=4k_F$) where the discretized branch is available only on the BEC side of the resonance.  For such values of ${\bf k}$ the calculation of the dynamic response factor can be usefully applied, within a LDA procedure, to estimate the response of the system in a trapped configuration. The results presented in these figures are based on a time-dependent generalization of the BCS mean field theory. This theoretical approach accounts for both phononic and single-particle parts of the excitation spectrum as well as for the corresponding hybridization phenomena. On the BEC side of the resonance one clearly sees a discretized peak correponding to the free molecule excitation energy $\hbar^2k^2/4m$. It is remarkable to see that even at unitarity, where  moleculae do not exist as independent excitations and the discretized peak has merged into the continuum of single particle excitations, the dynamic structure factor exhibits  a pronounced peak at $\sim \hbar^2 k^2/4m$. On the BCS side of the resonance  the molecular signatures are instead completely lost and the response is very similar to the one of an ideal Fermi gas. In the same figure we also show the magnetic structure factor  which is obtained replacing  the density operator
$\rho_q= \rho_{\uparrow}+\rho_{\downarrow}$ with the spin density operator $\rho_{\uparrow}-\rho_{\downarrow}$  in eq.(\ref{Sqomega}). The magnetic operator does not excite the phonon mode and its strength is restricted to the continuum of single particle excitations.

From the knowledge of the dynamic structure factor one can evaluate the  static structure factor, given by its frequency integral 
\begin{equation}
S(k)={\hbar\over N}\int_0^{\infty}S({\bf k},\omega)d\omega = {1\over N} <\sum_{i,j}e^{i{\bf k \cdot}({\bf r}_i-{\bf r}_j)}>
\label{staticfactor}
\end{equation}
where, in deriving the last equality, we have used definition (\ref{Sqomega}) and the completeness relationship
$\sum_n|n><n|=1$.
The static structure factor is related to the two-body correlation function $g(r)$ by the relationship
\begin{equation}
S(k)= 1+n\int d{\bf r}[g(r)-1]e^{-i{\bf k \cdot r}} 
\label{Sg2}
\end{equation}
Its behavior, evaluated at different points along the BEC-BCS crossover, is shown in Fig.13 where we report both the calculation of  the dynamic mean field approach \cite{combescot06}, which has been obtained by directly integrating the dynamic structure factor, and the results of the {\it ab initio} Monte Carlo calculations of \cite{stefano},  obtained by Fourier transforming the pair correlation function $g(r)$. The static structure factor decreases linearly at small $k$ as a consequence of the phononic nature of the excitation spectrum, while it approaches the incoherent value $S(k)=1$ for large wave vectors where only the terms $i=j$ in the sum of eq.(\ref{staticfactor}) survive. It worth noticing that at intermediate values of $k$ the static structure factor exhibits a maximum   which, in the BEC regime $k_Fa<<1$, takes the value $S(k)=2$ over an extended region of values of $k$. The origin of this plateau is directlly related to the molecular nature of elementary excitations. In fact,  if $k$ is larger than the Fermi momentum, but still small compared to the inverse of the size $a$ of the molecules, the probe mainly excites free molecules with energy $\hbar^2 k^2/4m$. Using the model independent f-sum rule result
\begin{equation} 
\hbar^2 \int d\omega S(k,\omega)\omega = N{\hbar^2 k^2 \over 2m}
\label{frule}
\end{equation}
and assuming, following Feynman, that a single molecular-like excitation with energy $\hbar^2k^2/4m$ exhausts the integral, one finds the value $S(k)=2$. Although this result holds only in the deep BEC limit, the figure shows that the enhancement of $S(k)$ with respect to the incoherent atomic value $S(k)=1$ is clearly visible also at unitarity.

\section{Rotating Fermi gases and superfluidity}

Superfluidity shows up in spectactular rotational features. In fact a superfluid cannot rotate like a rigid body, due to the irrotationality constraint (\ref{vphase}) imposed by the existence of the order parameter. At low angular velocity an important macroscopic  consequence of irrotationality is the quenching of the moment of inertia.  At higher angular velocities the superfluid can instead carry angular momentum via the formation of singular vortex lines. The circulation of these lines is quantized. In the presence of many vortex lines a regular vortex lattice is formed and the angular momentum acquired by the system takes the classical rigid body value. Both the quenching of the moment of inertia and the formation of vortex lines have been the object of fundamental investigation in the physics of quantum liquids and have been more recently explored also in  dilute Bose-Einstein condensed gases. In this section we summarize some of the main features exhibited by dilute Fermi gases where important experimental results are already available. We first  discuss the macroscopic consequences of the irrotationality constraint (moment of inertia and collective oscillations) and then  some key  features of quantized vortices. 

The moment of inertia $\Theta$ relative to the $z-th$ axis is defined as the response of the system to a rotational field $-\Omega L_z$ according to the relationship
\begin{equation}
<L_z> = \Omega \Theta
\label{Theta}
\end{equation}
where $L_z$ is the third component of  angular momentum  and the average is taken on the stationary configuration in the presence of the perturbation. In the limit of small angular velocity one can employ the formalism of linear response theory and write the moment of inertia in the form 
\begin{equation}
\Theta = 2Q^{-1}\sum_{m,n}e^{-\beta E_M}{|<n|L_z|m>|^2 \over E_n-E_m}
\label{Thetapert}
\end{equation}
where $|n>$ and $E_n$ are the eigenstates and eigenenergies of the unperturbed Hamiltonian and $Q$ is the partition function. There
is a simple case where the sum (\ref{Thetapert}) can be calculated explicitly. This is the ideal gas trapped by a deformed harmonic potential where the moment of inertia takes the analytic form \cite{sandro96a}
\begin{eqnarray}
\Theta &=&{mN  \over \omega^2_x-\omega^2_y} [<y^2>-<x^2>)(\omega^2_x+\omega^2_y) \nonumber \\
 &+& 2(\omega^2_y<y^2>-\omega^2_x
<x^2>)]
\label{ThetaS}
\end{eqnarray}
Result (\ref{ThetaS}) holds for both the Bose and Fermi ideal gas. It assumes $\omega_x\neq \omega_y$, but admits a well defined limit when $\omega_x\to \omega_y $. In the Fermi case, when the number of particles is large, one can use the semiclasssical estimate for the radii yielding $<x^2> \propto 1/\omega^2_x$ and $<y^2> \propto 1/\omega_x^2$. In this case eq.(\ref{ThetaS}) reduces  to the rigid value of the moment of inertia:
\begin{equation}
\Theta_{rig}=Nm<x^2+y^2>
\label{Thetarig}
\end{equation}
(for a non interacting Bose-Einstein condensed gas, at $T=0$, where the radii scale according to  $<x^2> \propto 1/\omega_x$ and $<y^2> \propto 1/\omega_x$, one instead finds that $\Theta \to 0$ as  
$\omega_x\to \omega_y$). 

Interactions can change the value of the moment of inertia of  a Fermi gas in a profound way. The simplest way to calculate $\Theta$ in the superfluid phase is to use the irrotational hydrodynamic equations developed in the previous section, by considering a trap rotating with angular velocity $\Omega$ and looking for the stationary solutions in the rotating frame. The resulting value for the angular momentum $<L_z>$ then permits to evaluate the moment of inertia through definition (\ref{Theta}). A similar procedure was implemented experimentally to generate the rotation of a dilute Bose gas \cite{dalibard1}

The equations of motion in the frame rotating with the trap are easily obtained by adding the term $-\Omega L_z$ to the Hamiltonian. The hydrodynamic equations, in the rotating frame, take the form
\begin{equation}
{\partial \over \partial t}n + \nabla {\bf \cdot} (n ({\bf v}-{\bf \Omega \times r})  =0
\label{continuity}
\end{equation}
 and  
\begin{equation}
m{\partial \over \partial t}{\bf v} + \nabla \left({1\over 2}m{\bf v}^2 + \mu(n) + V_{ho} - m{\bf v \cdot(\Omega \times r}) \right)  =0
\label{euler7}
\end{equation}
where the rotating trap is now described by the time independent potential $V_{ho}$. One sees that the rotation affects both the equation of continuity and the Euler equation. Here ${\bf v}= (\hbar/2m)\nabla \Phi$, where $\Phi$ is the phase of the order parameter, is the superfluid velocity in the laboratory frame, expressed in terms of the coordinates of the rotating frame. In the presence of harmonic trapping
an important class of stationary solutions is obtained making the ansatz
\begin{equation}
{\bf v}=\alpha \nabla(xy) \; .
\label{valpha}
\end{equation}
The equilibrium density, derivable from eq.(\ref{euler7}) by setting $\partial{\bf v}/\partial t=0$, has the same Thomas-Fermi form as in the absence of rotation, with renormalized values of the oscillator frequencies:
\begin{eqnarray}
\tilde{\omega}^2_x&=&\omega^2_x +\alpha^2 -2\alpha \Omega \nonumber \\
\tilde{\omega}^2_y&=&\omega^2_y +\alpha^2 +2\alpha \Omega 
\label{omegaalpha}
\end{eqnarray}
while the equation of continuity yields the relationship
\begin{equation}
\alpha = -\delta \Omega
\label{alphaomega}
\end{equation}
where 
\begin{equation}
\delta = {<y^2-x^2> \over <y^2+x^2>}
\label{delta}
\end{equation}
is the deformation of the atomic cloud in the $x-y$ plane. For small angular velocities one finds $\delta=\epsilon$ where 
\begin{equation}
\epsilon={\omega^2_x-\omega^2_y\over \omega^2_x+\omega^2_y}
\label{epsilon}
\end{equation}
is the deformation of the trap. By evaluating the angular momentum $L_z= m\int d{\bf r} ({\bf r \times v})n$ with the velocity field (\ref{valpha}) one finally finds that the moment of inertia is given by the irrotational form 
\begin{equation}
\Theta = \delta^2 \Theta_{rig}
\label{Thetairrot}
\end{equation}
which vanishes for an axi-symmetric configuration, pointing out the crucial role  played by superfluidity in the rotation of the gas. 

The measurement of the moment of inertia is not directly  accessible in dilute gases. However useful information on the rotational properties of the system can be obtained through the study of the collective oscillations. 
In this context a special role is played by the so called scissors 
mode, an oscillation of the system caused by the sudden rotation of a deformed trap. If the angle of rotation $\phi_0$ of the trap is small compared to the deformation of the trap (\ref{epsilon}) the rotation of the potential (\ref{Vho}) produces the perturbation $\delta V_{pert}=(\omega^2_x-\omega^2_y)\phi_0xy$ which naturally excites the quadrupole mode. The behavior of the resulting oscillation depends in a crucial way on whether the system is normal or superfluid. In fact, while the restoring force is, in both cases, proportional to the square of the deformation parameter of the trapping potential, the mass parameter, being proportional to the moment of inertia, behaves quite differently in the two cases.  In the superfluid the problem can be addressed by solving   the hydrodynamic equations with the ansatz $\delta n \propto xy$ and ${\bf v}\propto \nabla xy$. One easily finds that the oscillation around  equilibrium is characterized by the frequency \cite{scissors}
\begin{equation}
\omega = \sqrt2 \omega_\perp
\label{scissors}
\end{equation}
where $\omega^2_\perp= \omega^2_x+\omega^2_y$.
The result holds also for a tri-axially deformed trap where $\omega_x\neq \omega_y\neq \omega_z$.   For a  normal gas in the collisionless regime one  finds two frequencies at $\omega_{\pm}=|\omega_x\pm\omega_y|$. For a normal gas in the collisional regime one instead predicts an   oscillation with the same frequency (\ref{scissors}) in addition to a low frequency mode of diffusive nature, caused by the viscosity of the fluid. The different behavior exhibited by the normal gas (both in the collisionless and collisional phases) reflects the rigid nature of the classical moment of inertia.  The scissors mode, previously observed in a Bose-Einstein condensed gas \cite{foot}, has been recently investigated also in  ultracold Fermi gases \cite{grimmprivate}. At unitarity the experiment has confirmed the correctness of the hydrodynamic frequency (\ref{scissors}), while deeply in the BCS regime the beating between the frequencies $\omega_{\pm}=|\omega_x\pm\omega_y|$ has revealed the transition to the collisionless regime. If the gas is normal, but too deeply in the collisional hydrodynamic regime,  the diffusive mode predicted by classical hydrodynamics would be located at too low frequencies to be observable in experiments and the scissors mode would look the same as in the superfluid phase. This is expected to be the case at unitarity just above the critical temperature, thereby making the distinction between the superfluid and the normal phase, based on the study of the scissors mode, a difficult task.

More promising perspectives to distinguish between superfluid and collisional hydrodynamics are provided by the study of the collective oscillations excited on top of a rotating configuration.  In fact in the presence of vorticity $\nabla {\times v} \neq 0$ the equations of collisonal hydrodynamics contain an addiditional term depending on the curl of the velocity field, which is absent in the irrotational equations of superfluid hydrodynamics. In the laboratory frame the equation for the velocity field given by collisional hydrodynamics is given by 
\begin{eqnarray}
&m&{\partial \over \partial t}{\bf v} + \nabla \left({1\over 2}m{\bf v}^2 + \mu(n) + V_{ho}\right) 
-m{\bf v \times}({\bf \nabla \times v})  =0
\label{eulerR}
\end{eqnarray}
where we have omitted viscosity effects. The  collective oscilations correspond to linearized solutions with
$n=n_0+ \delta n$ and ${\bf v}= {\bf v}_0+\delta{\bf v}$. 
In the presence of a steady rotation of the trap at angular velocity $\Omega$ the oscillation frequencies resulting from the equations of rotational hydrodynamics   differ from the ones of superfluid hydrodynamics since in the former case the steady velocity field ${\bf v}_0$ is given   by the rigid body value ${\bf \Omega\times r}$, while in the latter case is given by the irrotational value $-\Omega\delta\nabla {\bf xy}$.  
After generating the steady rotation of the gas by adiabatically ramping the angular velocity of a deformed trap, one can suddenly stop the rotation and the system starts oscillating performing rotations around the symmetry axis of the trap. In the superfluid this procedure will excite the  scissors mode with frequency $\sqrt2 \omega_\perp$. In the case of rotational hydrodynamcis the behavior will instead be different, the scissors mode being coupled with the rigid rotation of the cloud. Under the condition $\Omega \gg \epsilon^2\omega_\perp$ the resulting oscillation is characterized by the beating law  $\phi(t)=(\Omega/\sqrt2 \omega_\perp)sin(\sqrt2 \omega_\perp t)cos(\Omega t)$ \cite{cozzini}. 
 
The possibility of distinguishing between superfluid and rotational hydrodynamics is a unique opportunity provided by ultracold Fermi gases. In fact  Bose-Einstein condensed gases, above $T_c$, are usually  extremely dilute and collisionless. Viceversa   in the Fermi case the normal gas can be easily dominated by collisions even at temperatures smaller than the Fermi energy as proven  by the behavior of the aspect ratio during the expansion  of the  unitary gas (see the previous Section).  

Let us now discuss the behavior of the superfluid Fermi gas at higher angular velocities where quantized vortices are formed.  Recent experiments    have confirmed  their existence along the BEC-BCS crossover (see Fig.14). In these experiments vortices  are produced by spinning the atomic cloud with a laser beam and are   observed after expansion  
by ramping the value of the scattering length to positve values in order to increase their visibility. Quantized vortices were previously extensively investigated with Bose-Einstein condensed gases  \cite{dalibard}. Quantized vortices  emerge as stable  
configurations  if the angular velocity  exceeds a critical value fixed by the energy cost needed for their production.  

A quantized vortex along the $z$ axis is associated with the appearence of a phase in the order parameter  (\ref{orderparameter}) given by the form $\exp{i\phi}$ where $\phi$ is the  angle around the $z-$-axis.
This yields the complex form 
\begin{equation}
\Delta({\bf r})= \Delta(r_\perp,z)\exp(i\phi)
\label{deltaphi}
\end{equation}
for the order parameter  $\Delta$ where, for simplicity, we have assumed that the system exhibits axial symmetry and we have used cylindrical coordinates. The velocity field ${\bf v}=(\hbar /2m)\nabla \phi$ of the vortex configuration has a tangential form with  modulus 
\begin{equation}
v= {\hbar \over 2m r_\perp}
\label{vv}
\end{equation}
which increases as one approaches the vortex line, in contrast to the rigid body dependence ${\bf v}={\bf \Omega \times r}$ characterizing the rotation of a normal fluid. 
The circulation is quantized according to the rule
\begin{equation}
\oint {\bf v \cdot} d{\bf \ell} = {\pi \hbar \over m}
\label{circulation}
\end{equation}
which is smaller by a factor $2$ with respect to the case of a Bose superfluid with the same atomic mass. The value of the circulation is independent of the radius of the contour. This is a consequence of the fact that the vorticity is concentrated on the $z$-axis according to the law
\begin{equation}
\nabla {\bf \times v}= \pi  {\hbar \over m} \delta^{(2)}({\bf r}_\perp)\hat{{\bf z}}
\label{vorticity}
\end{equation}
where $\hat{{\bf z}}$ is the unit vector in the z-th direction. It deeply differs 
from the uniform vorticity  $\nabla {\bf \times v}=2{\bf \Omega}$ of the rigid body rotation.

The angular momentum carried by the vortex is  given by the expression
\begin{equation}
L_z = m\int d{\bf r} {\bf r \times v} n({\bf r})= N{\hbar \over 2} 
\label{Lv}
\end{equation}
holding if the vortex line coincides with the symmetry axis of the density profile.  
If the vortex is displaced towards the perifery of a trapped gas the angular momentum takes a smaller value. In this case the axial symmetry of the problem is lost and the order parameter cannot be written in the form (\ref{deltaphi}).

A first estimate of the  energy of the vortex line is obtained employing macroscopic arguments based on hydrodynamics and considering, for simplicity, a gas confined in cylinder of radial size $R$. The energy $E_v$ acquired by the vortex   is   mainly determined by the hydrodynamic kinetic energy $(m/2)\int d{\bf r}v^2n$ which, employing the velocity field (\ref{vv}), yields the following estimate
for the vortex energy:
\begin{equation}
E_v = N{\hbar \over 4m R^2}ln{R\over \xi}
\label{EvM}
\end{equation}
where we have introduced the radius $\xi$ of the core of the vortex which fixes the distance below which the hydrodynamic expression for the kinetic energy no longer applies. This size is identified with the healing length whose value   varies significantly along the BEC-BCS crossover, being particularly large in the BCS limit.  Equation (\ref{EvM}) can be used to evaluate the critical angular velocity $\Omega_c$ for the existence of an energetically stable vortex line. This value is obtained by imposing that the change in the energy $E - \Omega_c L_z$ acquired by the system in the frame rotating with  angular velocity $\Omega_c$ be equal  to $E_v$. One finds
\begin{equation}
\Omega_c = {\hbar \over 2m R^2}ln{R\over \xi}
\label{omegac}
\end{equation}
Applying this estimate  to a harmonically trapped configuration with $R_{TF}\sim R$ and neglecting the logaritmic term which provides only a correction of order of unity, we find  $\Omega_c/\omega_\perp \simeq \hbar\omega_\perp /E_{ho}$ where $\omega_\perp$ is the radial  frequency of the harmonic potential and $E_{ho} \sim m\omega^2_\perp R^2_{TF}$ is the harmonic oscillator energy of the trapped gas. The above estimate shows that in the Thomas-Fermi regime, where $E_{ho}\gg \hbar \omega_{\perp}$, the critical frequency is much smaller than the radial trapping frequency, thereby  suggesting that vortices should be easily produced in slowing rotating traps. This conclusion, however, does not take into account
the fact that the nucleation of vortices  is strongly inhibited at low angular velocities by the occurrence of a barrier. For example in rotating Bose-Einstein condensates it has been experimentally shown that it is possible to  increase  the angular velocity of the trap up to values significantly  higher than $\Omega_c$ without generating vortical states. Under these conditions the response of the superfluid is governed by the equations of irrotational hydrodynamics.

A challenging problem concerns the visibility of the vortex lines. Due to the smallness of the healing length they cannot be observed {\it in situ},  but only after expansion. In particular the healing length is very small in the most interesting  unitary regime.   Another  difficulty in revealing  the vortex lines is the reduced contrast in the density with respect to the case of Bose-Einstein condensed gases. Actually, while the order parameter vanishes on the vortex line the density does not, unless one works in the deep BEC regime. Microscopic calculations of the vortex structure along the BEC-BCS crossover \cite{vortexmicro}  are in most cases based on the generalization of the BCS mean field theory to include non uniform configurations.  In Fig.15 we report the predictions obtained in \cite{vortexmicro} for the density profile  and for the order parameter $\Delta$ as a function of the distance from the vortical line.   The figure clearly shows that the contrast in the density profile  becomes weaker and weaker as one approaches the BCS regime where, differently from the opposite BEC regime, the density is not affected by the presence of the vortex. Conversely the order parameter $\Delta$ always vanishes close to the vortex line. 
   
At higher angular velocities more vortices can be formed giving rise to  a regular  vortex lattice. In this limit  the angular momentum acquired by the system takes the classical rigid body value and the rotation  will look similar to the one of a rigid body, characterized by the law $\nabla {\bf \times  v}= 2 \Omega$. Using result 
(\ref{vorticity}) and averaging the vorticity over several vortex lines one finds the result  $\nabla {\bf \times  v}= (h/2m) n_v \hat{{\bf z}}$, where $n_v$ is the number of vortices per unit area, so that the density of vortices is related to the angular velocity $\Omega$ by the useful relation 
\begin{equation}
n_v={2m \over \pi \hbar}\Omega
\label{densityV}
\end{equation}
which turns out to be a factor $2$ larger than in the case of Bose superfluids with the same value of the atomic mass. Equation (\ref{densityV}) shows that the distance between vortices (proportional to $1/\sqrt{n_v}$)  depends on the angular velocity but not on the density of the gas. In other words the vortices form a regular lattice even if the average density is not uniform as happens in the presence of harmonic trapping. This intriguing feature, already pointed out in the case of Bose-Einstein condensed gases, has been confirmed by the recent experiments on Fermi gases (see fig.14). 

The vortex lattice is responsible for an important bulge effect associated with the increase  of the radial size of the cloud.  In fact in the presence of a rigid rotation the effective potential felt by the atoms is given by $V_{ho}-(m/2) \Omega^2 r^2_\perp$, corresponding to a modified equilibrium density, whose Thomas-Fermi radii satisfy the new relationship
\begin{equation}
{R^2_z \over R^2_\perp} = {\omega^2_\perp-\Omega^2 \over \omega_z^2}
\label{bulge}
\end{equation}
showing, by the way, that at equilibrium the angular velocity cannot overcome the radial trapping frequency.
In experiments where the vortex lattice is not formed in the presence of a stationary  rotating  trap this formula  can be used to evaluate the effective value  of the angular velocity  by just measuring the aspect ratio. 

Important consequences of the vortex lines concern also the frequency of the collective oscillations. For example, using a sum rule approach \cite{zambelli} it is possible to show that the splitting between the $m=\pm 2$
  quadrupole frequencies  is given by the 
formula 
\begin{equation}
\omega(m=+2)-\omega(m=-2)=2{\frac{\ell_{z}}{m\langle r_{\perp
}^{2}\rangle }}  \label{33F}
\end{equation}
where $\ell_z=\langle L_{z}\rangle /N$ is the angular momentum per particle  carried by the vortical configuration. For a single vortex line $\ell_z$ is  equal to
$\hbar/2  $, while for a vortex lattice $\ell_z$ is given by the rigid body value $\Omega m <r^2_{\perp}>$. In the  latter case one recovers exactly the splitting $2\Omega$  predicted by the equations  of rotational hydrodynamics which, in the case of axi-symmetric configurations,  yield the result \cite{cozzini}
\begin{equation}
\omega(m=\pm2)=\sqrt{2\omega^2_{\perp}-\Omega^2}\pm\Omega
\label{m=2R}
\end{equation}
for the frequencies of the two $m=\pm 2$ quadrupole modes.
The equations  of rotational hydrodynamics actually provide the correct   description of the collective  oscillations of a gas containing a vortex lattice which, from a macroscopic point of view, behaves like a classical gas rotating in a rigid way.

The experimental production and measurement of quantized vortices  has provided a definitive proof of superfluidity in these ultracold Fermi gases. Actually the esixtence of vortices as stable configurations can be regarded as a proof of the absence of viscosity. In fact in the rotating frame the velocity field of a vortical configuration is not vanishing and any tiny presence of viscosity would bring the system into a rigid rotation, corresponding to a vanishing velocity in the rotating frame.
 
\subsection{Conclusions}

In this paper we have have summarized some features of the superfluid behavior exhibited by ultracold Fermi gases with special emphasis on the behavior of the collective oscillations.  Other subjects related to the dynamics of these novel systems that would deserve careful investigation concern  Fermi gases in the presence of inbalance ($N_\uparrow\ne N_\downarrow$) and Fermi gases of unequal masses ($m_\uparrow \ne m_\downarrow$).

It is a pleasure to acknowledge fruitful collaborations and discussions with the members of the CNR-INFM Center on Bose-Einstein Condensation in Trento. In particular  this paper has benefited from many discussions with Stefano Giorgini and Lev Pitaevskii. Stimulating collaborations with Roland Combescot are also acknowledged.

\section{QUESTIONS}

\begin{widetext}
\begin{figure}[ptb]
\begin{center}
\includegraphics[width=0.5\textwidth]{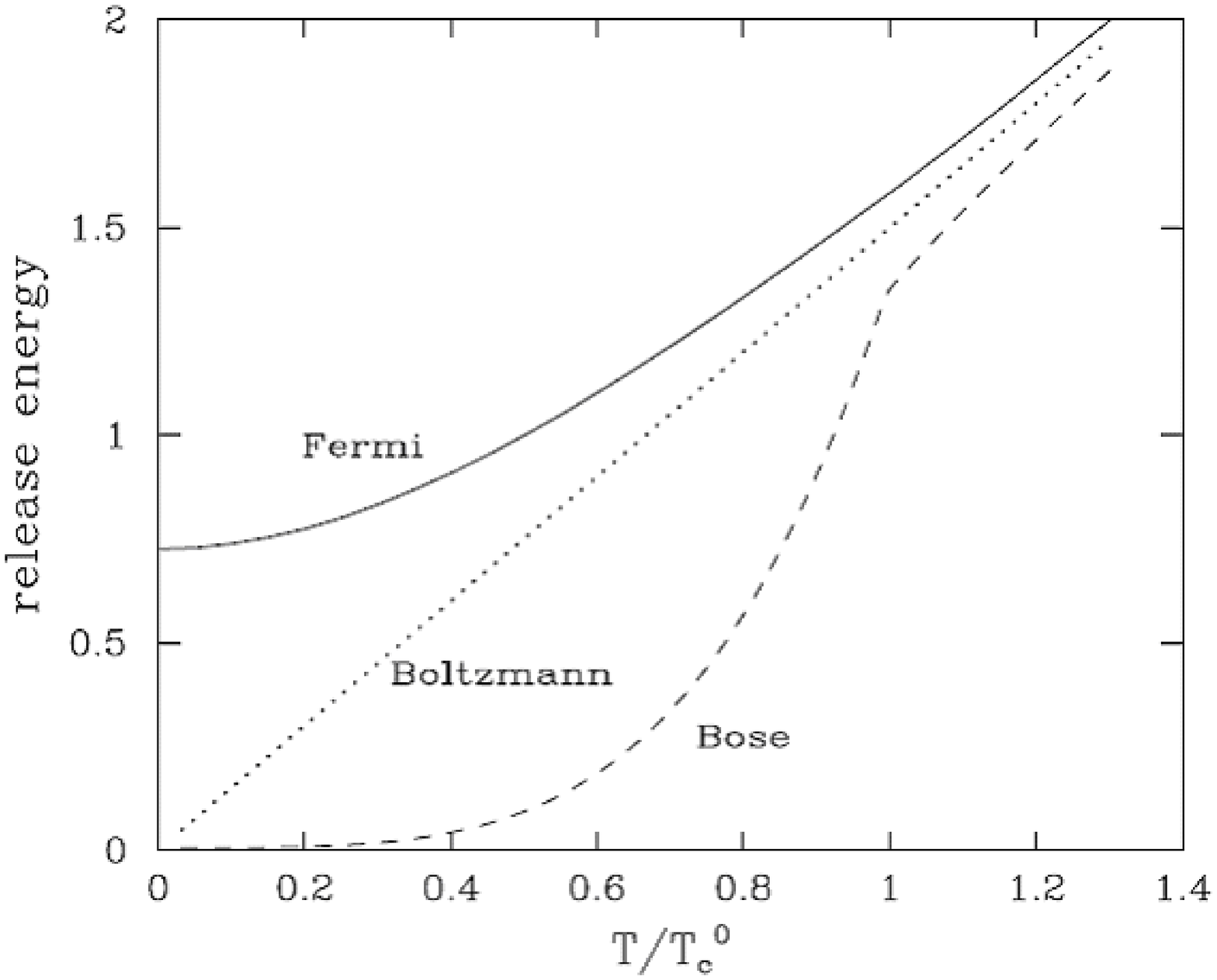}
\caption{Release energy in units of the Fermi energy (\ref{EF}) as a function of the temperature calculated for an ideal Fermi (solid line), classical (dotted line) and Bose (dashed line) gas with the same number of atoms and the same mass. Here $T^0_c$ is the critical temperature for Bose-Einstein condensation for an ideal Bose gas.} \label{Figure1}
\end{center}
\end{figure}

\begin{figure}[ptb]
\begin{center}
\includegraphics[width=0.5\textwidth]{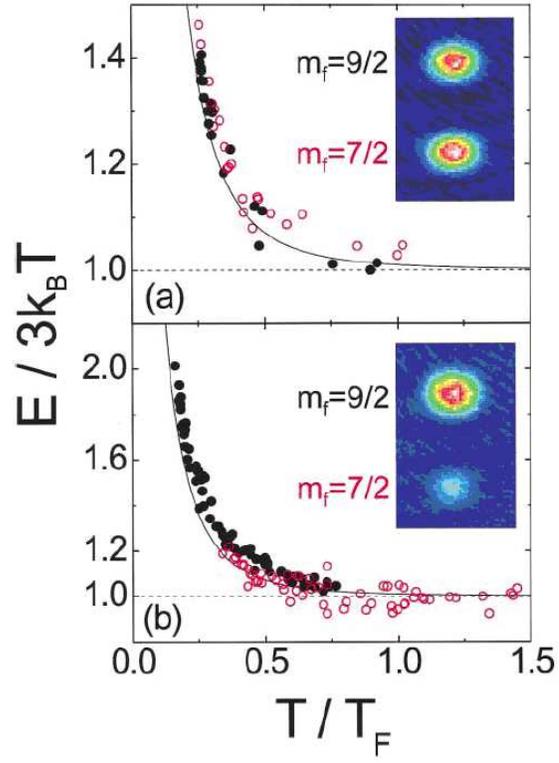}
\caption{Evidence for quantum degeneracy effects in trapped Fermi gases. The average energy per particle, extracted from absorption images, is shown for two spin mixtures. In the quantum degenerate regime the data agree well with the ideal Fermi gas prediction (solid line). From \cite{jinN}.} \label{Figure2}
\end{center}
\end{figure}

\begin{figure}[ptb]
\begin{center}
\includegraphics[width=0.5\textwidth]{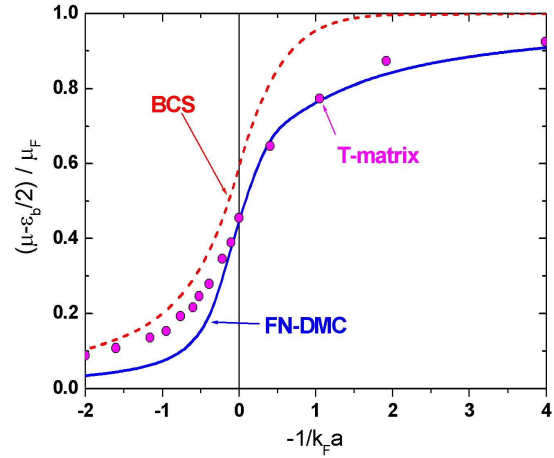}
\caption{Equation of state along the BEC-BCS crossover as a funtion of the dimensioneless parameter $1/k_Fa$. The results of the fixed node diffused Monte Carlo calculations (FN-DMC) of \cite{stefano} and of the diagrammatic expansion of \cite{strinatiT} are compared with the prediction of the BCS mean field theory. } \label{Figure3}
\end{center}
\end{figure}

\begin{figure}[ptb]
\begin{center}
\includegraphics[width=0.5\textwidth]{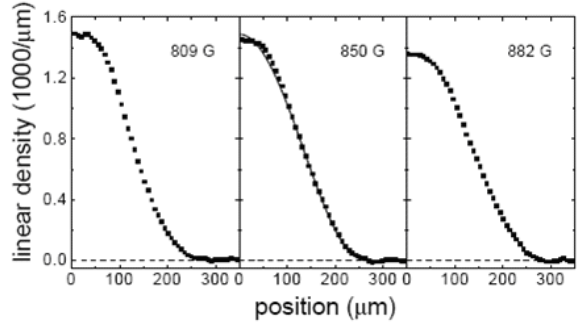}
\caption{Experimental results for the column density profiles along the BEC-BCS crossover for a gas of $^6{Li}$ atoms. The continuous  curve at $850 G$, corresponding to unitarity, is the best fit based on eq.(\ref{column}. From \cite{grimmdensity}).} \label{Figure4}
\end{center}
\end{figure}

\begin{figure}[ptb]
\begin{center}
\includegraphics[width=0.5\textwidth]{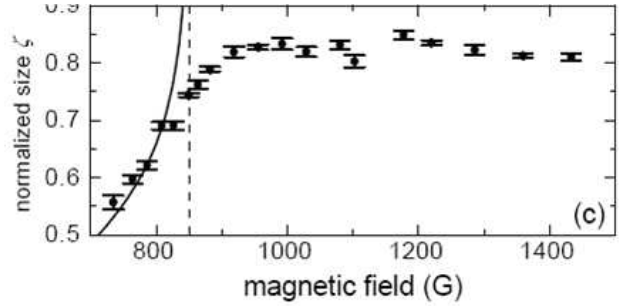}
\caption{Experimental results for the axial radius along the BEC-BCS crossover for a gas of $^6{Li}$ atoms. The data are plotted after normalization to the non interacting Fermi gas.  The full line corresponds to the prediction of Eq. (\ref{RBEC}) with $a_M=0.6 a$. From \cite{grimmdensity}).} \label{Figure5}
\end{center}
\end{figure}

\begin{figure}[ptb]
\begin{center}
\includegraphics[width=0.5\textwidth]{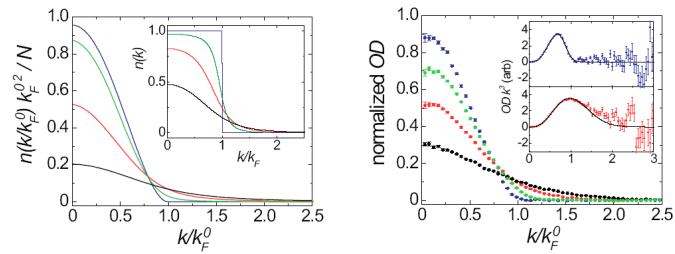}
\caption{Theoretical (left) vs experimental (right) values of the momentum distribution in a gas of $^{40}K$ atoms along the BEC-BCS crossover. From \cite{jin2}.} \label{Figure6}
\end{center}
\end{figure}

\begin{figure}[ptb]
\begin{center}
\includegraphics[width=0.5\textwidth]{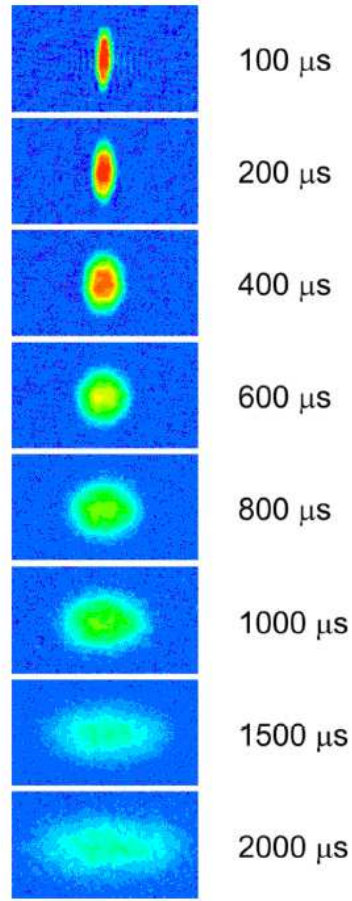}
\caption{Images of the expanding cloud of an ultracold Fermi gase of $^6Li$ atoms at different times. The figure clearly shows the typical inversion from the cigar to the disc shape revealing the hydrodynamic nature of the expansion. From \cite{ohara}. } \label{Figure7}
\end{center}
\end{figure}

\begin{figure}[ptb]
\begin{center}
\includegraphics[width=0.5\textwidth]{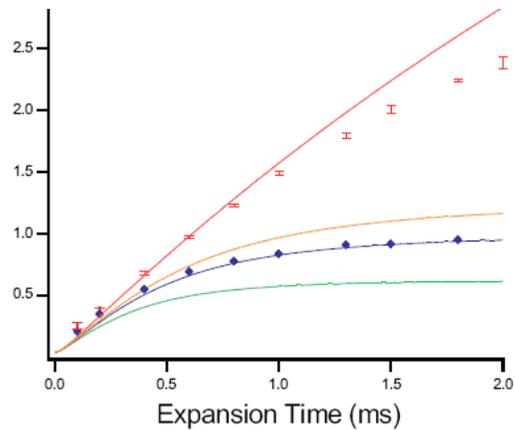}
\caption{Aspect ratio as a function of time during the expansion of an ultracold Fermi gas at unitarity (red points: experiment; red line: hydrodynamic theory). For comparison the figure also shows the results in the absence of interactions (blue points: experiment; blue line ballistic expansion). From \cite{ohara}.} \label{Figure8}
\end{center}
\end{figure}

\begin{figure}[ptb]
\begin{center}
\includegraphics[width=0.5\textwidth]{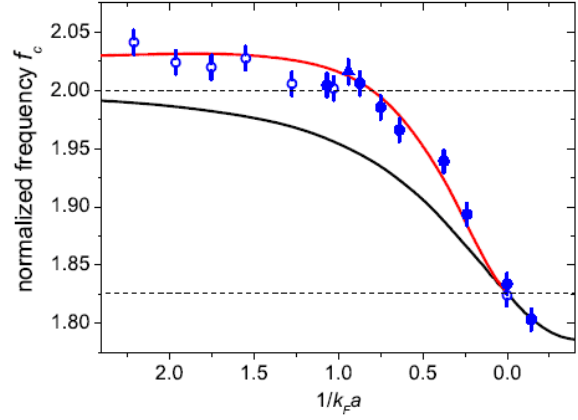}
\caption{Frequency of the radial compression mode for an elongated Fermi gas in units of the radial frequency. The theory curves refer to mean field BCS theory (lower curve) and quantum MC calculations (upper curve). From \cite{grimmoscillations}.
from \cite{grimmoscillations}. } \label{Figure9}
\end{center}
\end{figure}

\begin{widetext}
\begin{figure}[ptb]
\begin{center}
\includegraphics[width=0.3\textwidth]{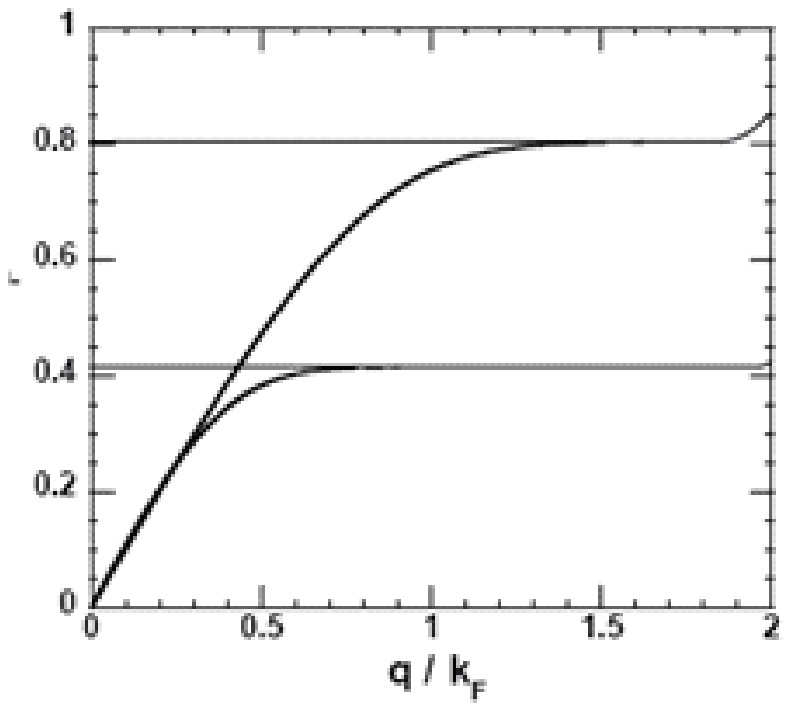}
\includegraphics[width=0.3\textwidth]{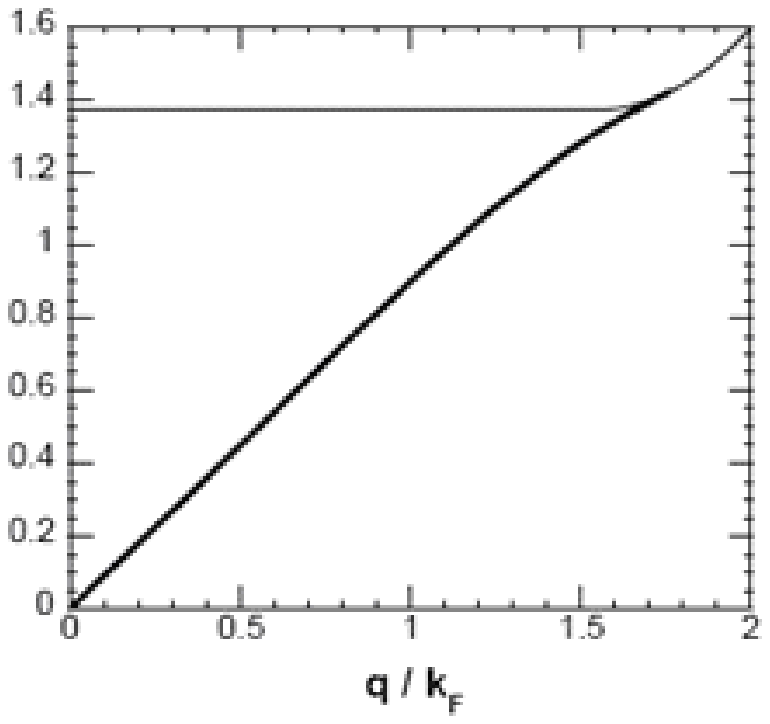}
\includegraphics[width=0.28\textwidth]{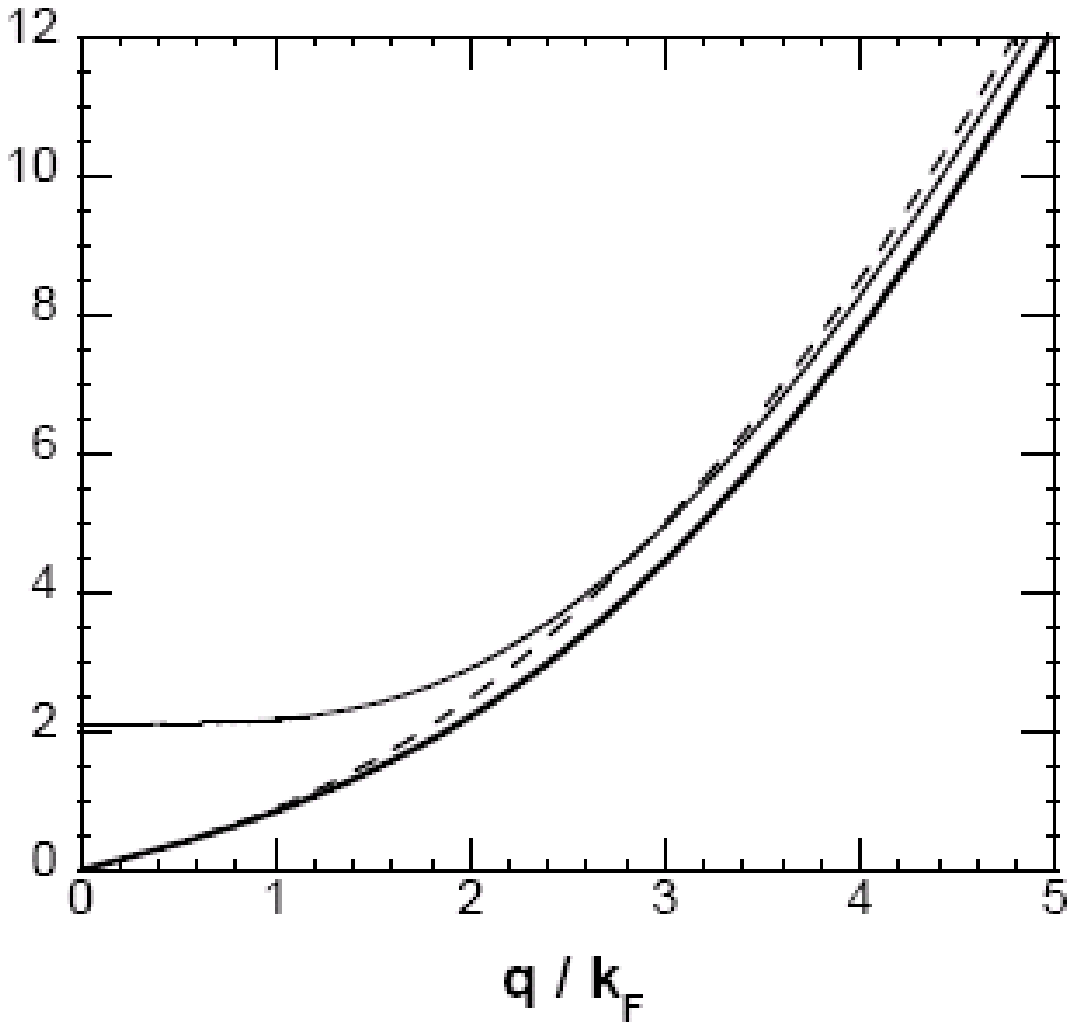}
\caption{Excitation spectrum of the superfluid Fermi gas along the BEC-BCS crossover. Energy is given in units of the Fermi energy. Left: BCS regime. Center: unitarity. Right: BEC regime. From \cite{combescotkagan}. } \label{Figure10}
\end{center}
\end{figure}
\end{widetext}

\begin{figure}[ptb]
\begin{center}
\includegraphics[width=0.5\textwidth]{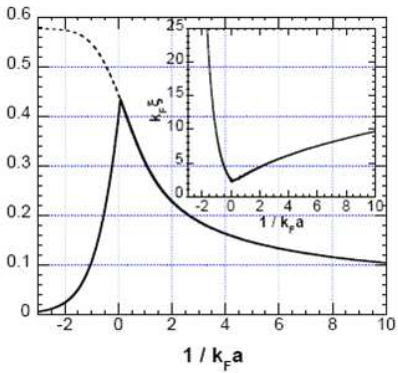}
\caption{Landau's critical velocity (in units of Fermi velocity) calculated along the crossover. The figure clearly shows that the critical velocity is largest close to unitarity. From \cite{combescotkagan}. } \label{Figure11}
\end{center}
\end{figure}

\begin{figure}[ptb]
\begin{center}
\includegraphics[width=0.5\textwidth]{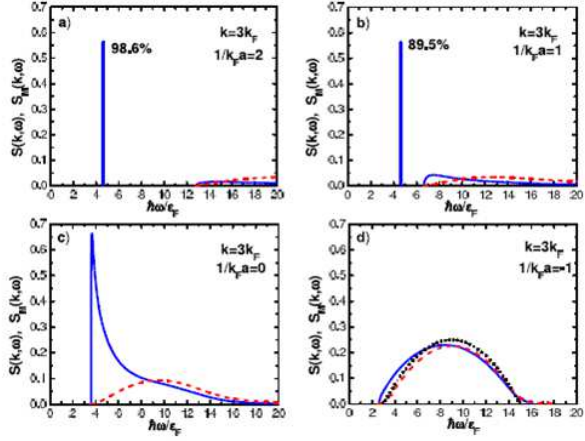}
\caption{Dynamic structure factor of the superfluid Fermi gas along the BEC-BCS crossover at $k=4k_F$. From  \cite{combescot06}.} \label{Figure12}
\end{center}
\end{figure}

\begin{figure}[ptb]
\begin{center}
\includegraphics[width=0.5\textwidth]{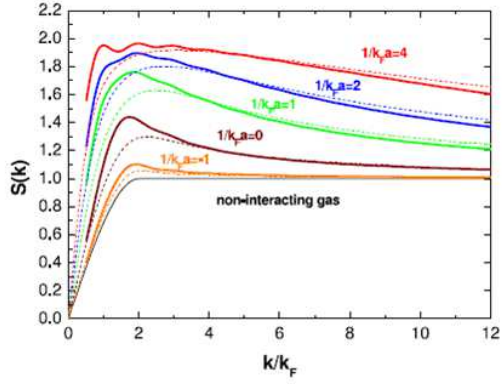}
\caption{Static structure factor of the superfluid Fermi gas along the BEC-BCS crossover. From \cite{combescot06}.} \label{Figure13}
\end{center}
\end{figure}

\begin{figure}[ptb]
\begin{center}
\includegraphics[width=0.5\textwidth]{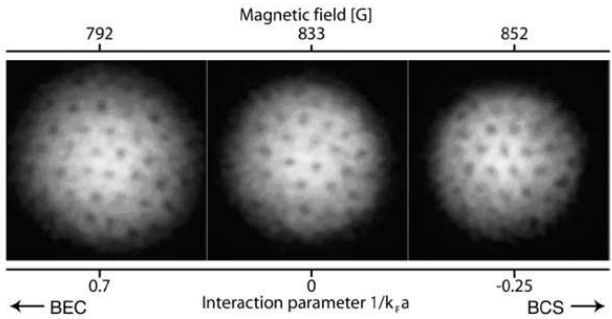}
\caption{Experimental observation of quantized vortices in a superfluid Fermi gas along the BEC-BCS crossover \cite{mitvortices}.} \label{Figure14}
\end{center}
\end{figure}

\begin{figure}[ptb]
\begin{center}
\includegraphics[width=0.5\textwidth]{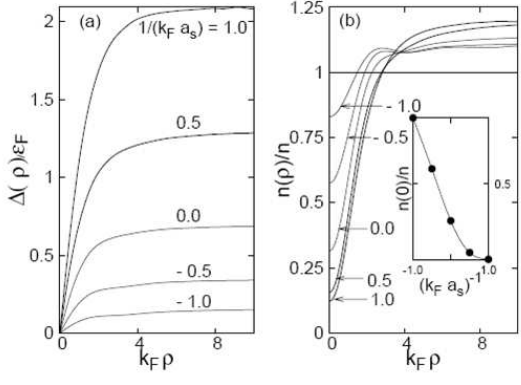}
\caption{Radial profiles of the density and of the order parameter of a vortex line along the BEC-BCS crossover. From R. Sensarma et al. \cite{vortexmicro}.} \label{Figure15}
\end{center}
\end{figure}
\end{widetext}

\end{document}